\documentclass[useAMS,usenatbib]{mn2e}

\usepackage{aas_macros,graphicx,amsmath}
\usepackage{color}
\usepackage{rotating}
\usepackage[labelsep=period,labelfont=bf]{caption}
\usepackage{lscape}
\usepackage{soul}

\title[Evolution in the size and luminosity of giant H{\sc ii} regions]{Hubble Space Telescope H$\alpha$ imaging of star-forming galaxies at $z \simeq 1 - 1.5$: evolution in the size and luminosity of giant H{\sc ii} regions}
\author[R.~C.~Livermore et~al.]{R.~C.~Livermore,$^{1}$ \thanks{E-mail:
r.c.livermore@durham.ac.uk}
T.~Jones,$^{2}$
J.~Richard,$^{1,3}$
R.~G.~Bower$^{1}$
R.~S.~Ellis,$^{2}$
\newauthor A.~M.~Swinbank,$^{1}$
J.~R.~Rigby,$^{4}$
Ian Smail,$^{1}$
H.~Ebeling$^{5}$
and R.~A.~Crain$^{6}$ \\
$^{1}$Institute for Computational Cosmology, Durham University, South Road, Durham DH1 3LE, UK\\
$^{2}$Astronomy Department, California Institute of Technology, 249-17, Pasadena, CA 91125, USA\\
$^{3}$CRAL Observatoire de Lyon, 9 Avenue Charles Andr\'e, 69561 Saint-Genis-Laval, France\\
$^{4}$NASA Goddard Space Flight Center, Code 665, Greenbelt, MD 20771, USA\\
$^{5}$Institute for Astronomy, University of Hawaii, 2680 Woodlawn Drive, HI 96822, USA\\
$^{6}$Leiden Observatory, Leiden University, PO Box 9513, 2300 RA Leiden, The Netherlands\\
}
\begin{document}

\date{}

\pagerange{\pageref{firstpage}--\pageref{lastpage}} \pubyear{2011}

\maketitle

\label{firstpage}

\begin{abstract}
We present \emph{HST}/WFC3 narrowband imaging of the H$\alpha$ emission in a sample of eight gravitationally-lensed galaxies at $z = 1 - 1.5$. The magnification caused by the foreground clusters enables us to obtain a median source plane spatial resolution of $360$pc, as well as providing magnifications in flux ranging from $\sim10\times$ to $\sim50\times$. This enables us to identify resolved star-forming H{\sc ii} regions at this epoch and therefore study their H$\alpha$ luminosity distributions for comparisons with equivalent samples at $z \sim 2$ and in the local Universe. We find evolution in the both luminosity and surface brightness of H{\sc ii} regions with redshift. The distribution of clump properties can be quantified with an H{\sc ii} region luminosity function, which can be fit by a power law with an exponential break at some cut-off, and we find that the cut-off evolves with redshift. We therefore conclude that `clumpy' galaxies are seen at high redshift because of the evolution of the cut-off mass; the galaxies themselves follow similar scaling relations to those at $z=0$, but their H{\sc ii} regions are larger and brighter and thus appear as clumps which dominate the morphology of the galaxy. A simple theoretical argument based on gas collapsing on scales of the Jeans mass in a marginally unstable disk shows that the clumpy morphologies of high-$z$ galaxies are driven by the competing effects of higher gas fractions causing perturbations on larger scales, partially compensated by higher epicyclic frequencies which stabilise the disk.
\end{abstract}


\section{Introduction}
\label{sec:intro}

\begin{figure*}
\includegraphics[width=174mm, angle=0]{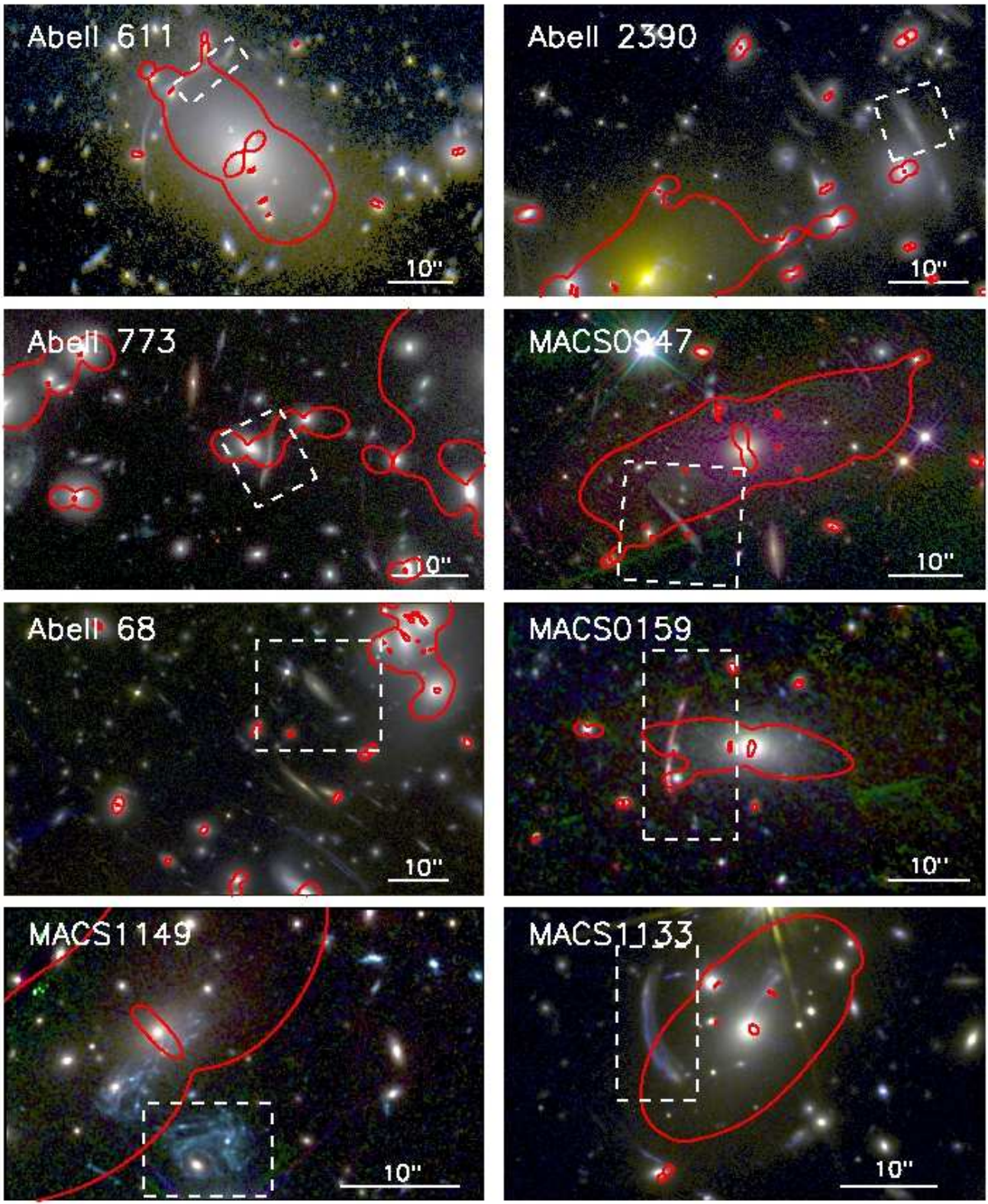}
\caption{HST/ACS and WFC3 three-colour images of the observed clusters with the critical line at the redshift of the target arc overlaid, showing the positions of the target arcs. The arcs are contained within the white dashed boxes which denote the regions extracted in Figure \ref{fig:Ha}.}
\label{fig:hst}
\end{figure*}

Observations of star-forming galaxies at high-$z$ have shown that a significant fraction of the population have turbulent, clumpy, rotating disks with clump masses of $\sim 10^{8-9}$M$_{\sun}$, a factor of $\sim100\times$ the typical Giant Molecular Cloud (GMC) locally \citep[e.g.][]{1995AJ....110.1576C,2004ApJ...612..191E,2005ApJ...627..632E,2009ApJ...706.1364F,2009ApJ...692...12E}. The clumps are thought to form from gravitational instabilities in gas-rich disks \citep{2007ApJ...658..763E,2009ApJ...692...12E,2008ApJ...687...59G,2010MNRAS.409.1088B}.

Some recent numerical simulations have suggested that the majority of massive, high-$z$ galaxies accrete their gas via `cold flows,' in which the gas is accreted smoothly along filaments. These cold flows are less disruptive than a major merger, and hence offer a route to maintain marginally stable disks (Toomre parameter $Q \sim 1$) without disrupting the structure and dynamics. Cold-flow accretion is expected to be a dominate mode of mass assembly above $z \simeq 1$, and thus accounts for the ubiquity of large clumps at high redshift \citep[e.g.][]{2009ApJ...694L.158B,2011ApJ...730....4B,2009ApJ...703..785D}.

In this picture, the clumps are considered to be transient features, forming in marginally unstable disks at high-$z$ and fed by smooth accretion of gas onto the galaxy. Clumpy galaxies therefore represent a phase in the evolution of present-day spiral disks.

There is a need to test the internal physical properties of the interstellar medium (ISM) observationally, to determine whether the clumps are scaled-up analogues of local H{\sc ii} regions or represent a different `mode' of star formation, and whether they can explain the strong evolution of star formation rate density with redshift. However, sufficient spatial resolution is required to resolve the ISM on the scales of star-forming regions. Even with the use of adaptive optics, spatially resolved studies of high-redshift galaxies to date have been limited to a resolution of $\sim 1.5$kpc \citep[e.g.][]{2006Natur.442..786G,2009ApJ...706.1364F}; using the \emph{Hubble Space Telescope (HST)}, only the largest starburst complexes can be resolved, on scales of $\sim 1$kpc \citep{2007ApJ...658..763E}. On these scales, it is possible to probe the dynamics of galaxies on large scales, and \citet{2011ApJ...733..101G} found evidence that $Q < 1$ in the regions of galaxies where clumps are found, lending observational support to the theory that the clumps form from internal gravitational instabilities. In order to study the clumps in detail, we need to resolve high-redshift disks on the scales of individual star-forming regions; in the local universe, this is $\sim 100$pc.

The required spatial resolution can currently only be achieved by exploiting gravitational lensing. By targeting galaxies that lie behind foreground cluster lenses, it is possible to benefit from linear magnification factors (along one direction) of up to $50\times$ \citep[e.g.][]{2007MNRAS.376..479S,2009MNRAS.400.1121S,2010MNRAS.404.1247J}, and to isolate H{\sc ii} regions of order $\sim 100$pc out to $z \sim 5$ \citep{2009MNRAS.400.1121S}. Regions were found with star formation surface densities $\Sigma_{\rmn{SFR}}$ $\sim 100\times$ higher than those found locally \citep{2009MNRAS.400.1121S,2010MNRAS.404.1247J}. These regions of dense star formation are comparable to the most intensely star-forming interacting systems in the local Universe \citep{2006A&A...445..471B}, yet appear to be ubiquitous in non-interacting galaxies at high redshift.

It is not known what drives these regions of intense star formation at high-$z$, although \citet{2010MNRAS.404.1247J} suggest a combination of higher gas density, increased star formation efficiency and shorter star-formation timescales. In addition, their data give the appearance of a bimodal distribution of H{\sc ii} region surface brightnesses, although there is no known physical process that might drive this. In order to understand this result further, we require a sample at intermediate redshift ($z \sim 1 - 1.5$) with which we can probe the evolution of star formation density with redshift at higher sensitivity so that regions comparable to those at $z = 0$ are detectable.

Previous work on high-$z$ clumps has made use of Integral Field Units such as \emph{Keck}/OSIRIS \citep{2010MNRAS.404.1247J,2012MNRAS.tmp.2831W}, \emph{Gemini}/NIFS \citep{2009MNRAS.400.1121S} and \emph{VLT}/SINFONI \citep{2009ApJ...706.1364F}. These allow detailed mapping of the nebular emission lines, but at lower sensitivity than is achievable with imaging. An alternative means of identifying star-forming regions with high sensitivity is to take imaging through narrowband filters. The Wide Field Camera 3 (WFC3) on the \emph{HST} presents an opportunity to study the star formation in galaxies at $z \sim 1$ and $z \sim 1.5$, as there are narrowband filters available which correspond to the wavelength of the H$\alpha$ emission line at these redshifts. Combining the sensitivity and high spatial resolution of \emph{HST}/WFC3 with the magnification afforded by gravitational lensing by foreground clusters, we can map the internal star formation distribution and so identify the frequency and properties of giant H{\sc ii} regions.

In this paper, we therefore study the star formation morphologies of eight galaxies at $z \sim 1 - 1.5$. We present the sample in \S \ref{sec:sample}, present the properties of the galaxies and their star-forming clumps in \S \ref{sec:results}, discuss the implications in \S \ref{sec:disc} and present our conclusions in \S \ref{sec:conc}. Throughout, we adopt a $\Lambda$CDM cosmology with $H_0 = 70$\,km\,s$^{-1}$\,Mpc$^{-1}$, $\Omega_{\Lambda} = 0.7$ and $\Omega_m = 0.3$. Star-formation rates are calculated from H$\alpha$ luminosity $L_{\rmn{H}\alpha}$ using the prescription of \citet{1998ARA&A..36..189K} adjusted to a \citet{2003PASP..115..763C} IMF.

\section{Sample and Observations}
\label{sec:sample}

Our sample comprises eight lensed galaxies, each with spectroscopically-confirmed redshifts between $1 < z < 1.5$ such that the H$\alpha$ emission line falls within the high-transmission region of the narrowband filters on WFC3. The associated cluster lenses are massive systems from the X-ray selected BCS and MACS samples \citep{1998MNRAS.301..881E,2001ApJ...553..668E,2007ApJ...661L..33E,2010MNRAS.407...83E} with well-constrained mass models (see references in Table \ref{tab:props}), so that the effects of lensing can be accounted for.

The positions and properties of the sample are given in Table \ref{tab:props}. We observed each target in the narrowband filter covering H$\alpha$ for a typical exposure time of 6\,ks (2 orbits), using a 3- or 4-point linear dithering pattern of $\pm$5 arcsecs in both directions to improve the detection and removal of cosmic ray hits and bad pixels. At the same time, three of the targets (MACS J0947, MACS J0159 and MACS J1133), which did not have WFC3 data in the archive, were observed in the corresponding broadband filter using the same sequence of observations as their corresponding narrowband data, for a total of 3\,ks (1 orbit). The narrowband data and new broadband observations were obtained in Cycle 18 under Program 12197 (PI:Richard), with the exception of Abell~2390, for which the broadband and narrowband data were taken in Cycle 17 under Program 11678 (PI:Rigby). The remaining broadband data were obtained under Cycle 17 Program 11591 (PI:Kneib) or Cycle 18 Program 12065-9 (PI:Postman) as indicated in the notes to Table \ref{tab:props}.

All of the WFC3 data were reduced using the multidrizzle software \citep{2002hstc.conf..337K} under \verb'PyRAF' to perform a cosmic-ray rejection, sky subtraction, and drizzling onto an output pixel scale of 0.05". The narrowband and broadband images of the same cluster were aligned using the location of $\sim 20$ bright stars. A narrowband excess image was constructed by direct pixel-to-pixel subtraction between the narrowband and broadband images, including an arbitrary scaling factor. We calibrated this scaling factor by checking that all bright cluster members, which are featureless elliptical galaxies with no emission lines in the respective filters, became consistent with the background in the excess image. For Abell~773 and Abell~68, the broadband images available in the archive did not directly overlap the H$\alpha$ emission line, so an estimate of the broadband continuum was made by linear interpolation between the adjacent F110W and F160W filters.

The flux calibration of each image was verified using {\sc 2MASS} stars in the fields, and in all cases was found to agree to within 15\%, which is sufficient precision for our purposes.

\begin{figure*}
\includegraphics[height=88mm, angle=90]{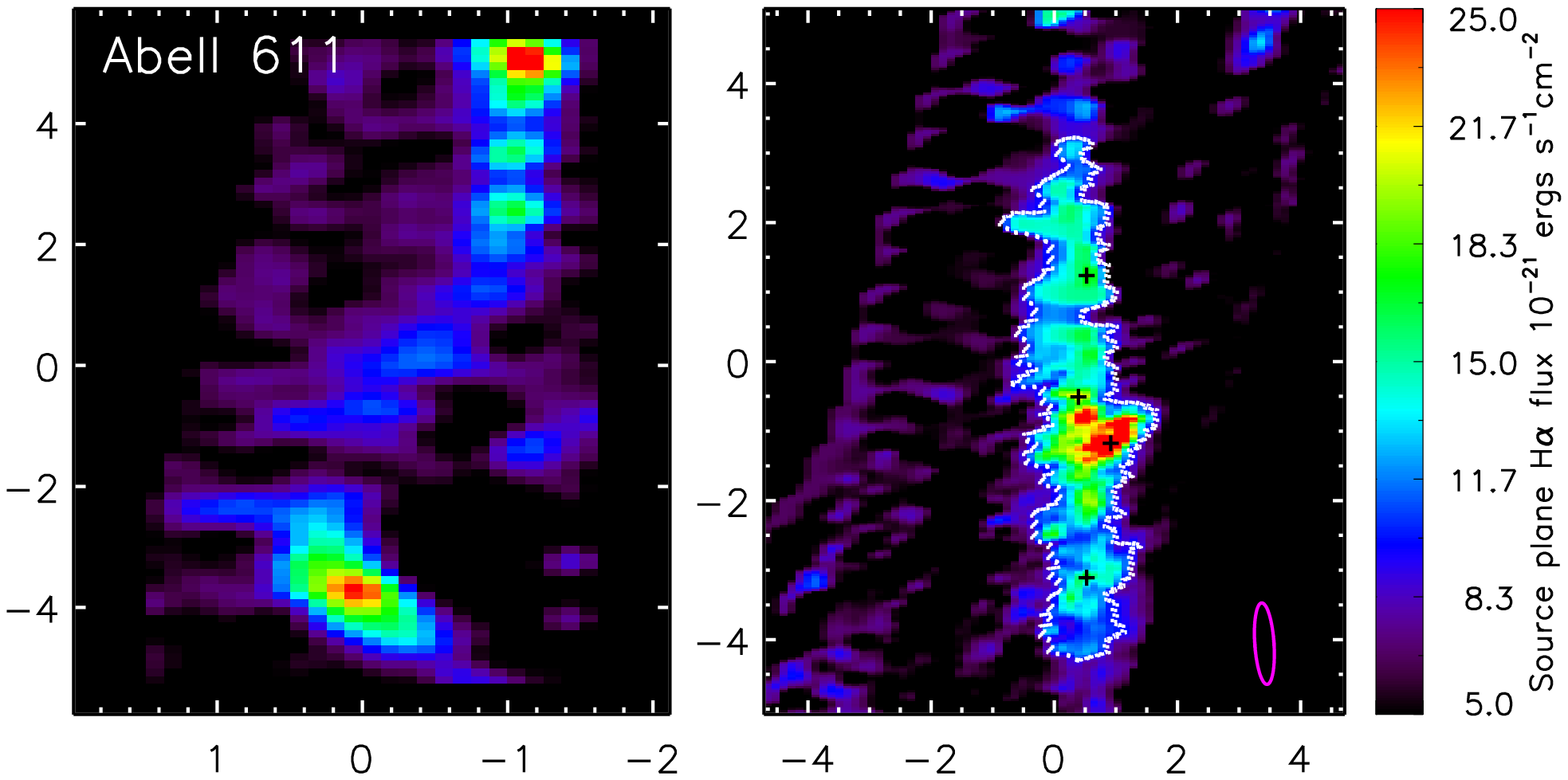}
\includegraphics[height=88mm, angle=90]{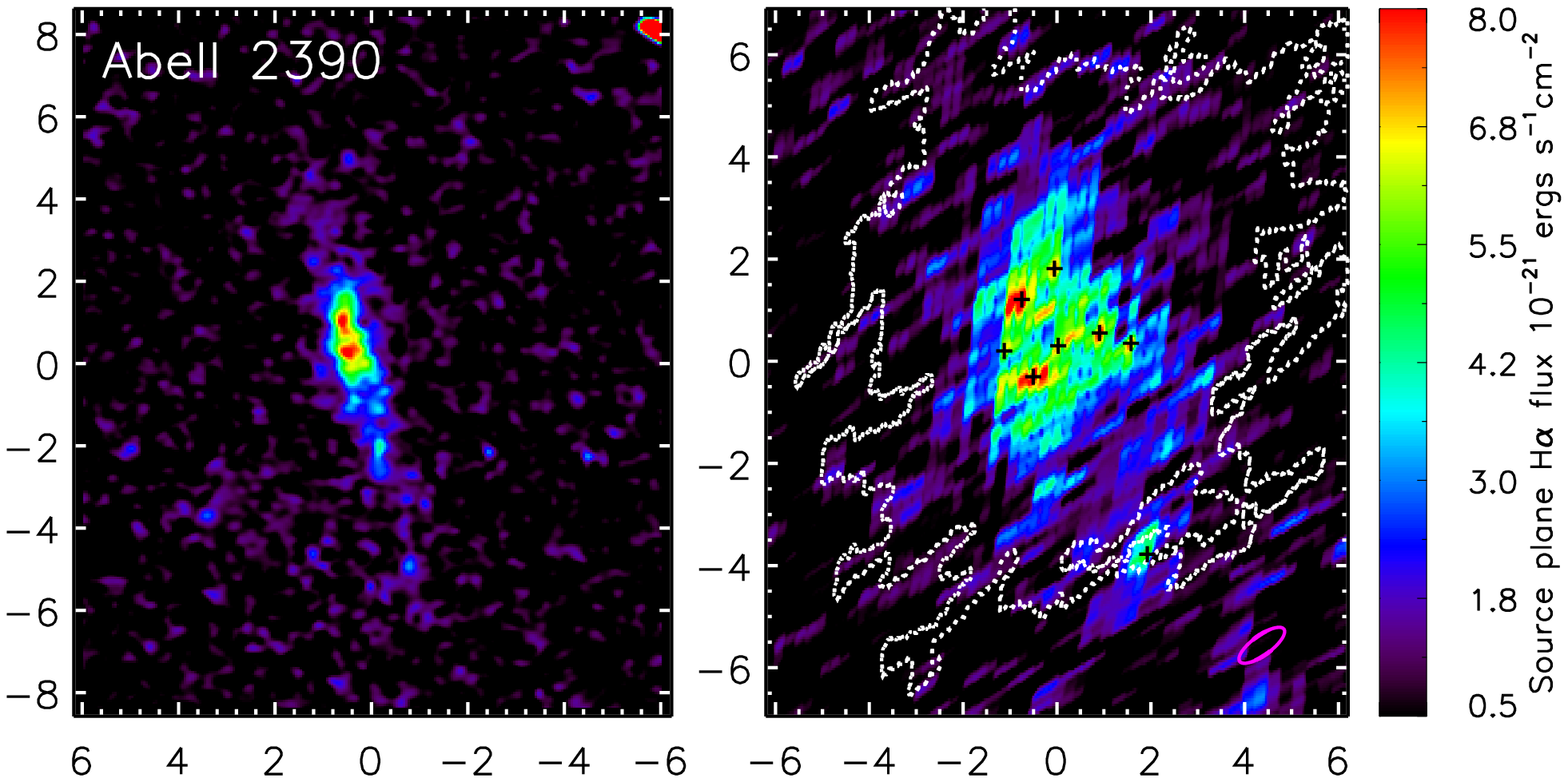}
\includegraphics[height=88mm, angle=90]{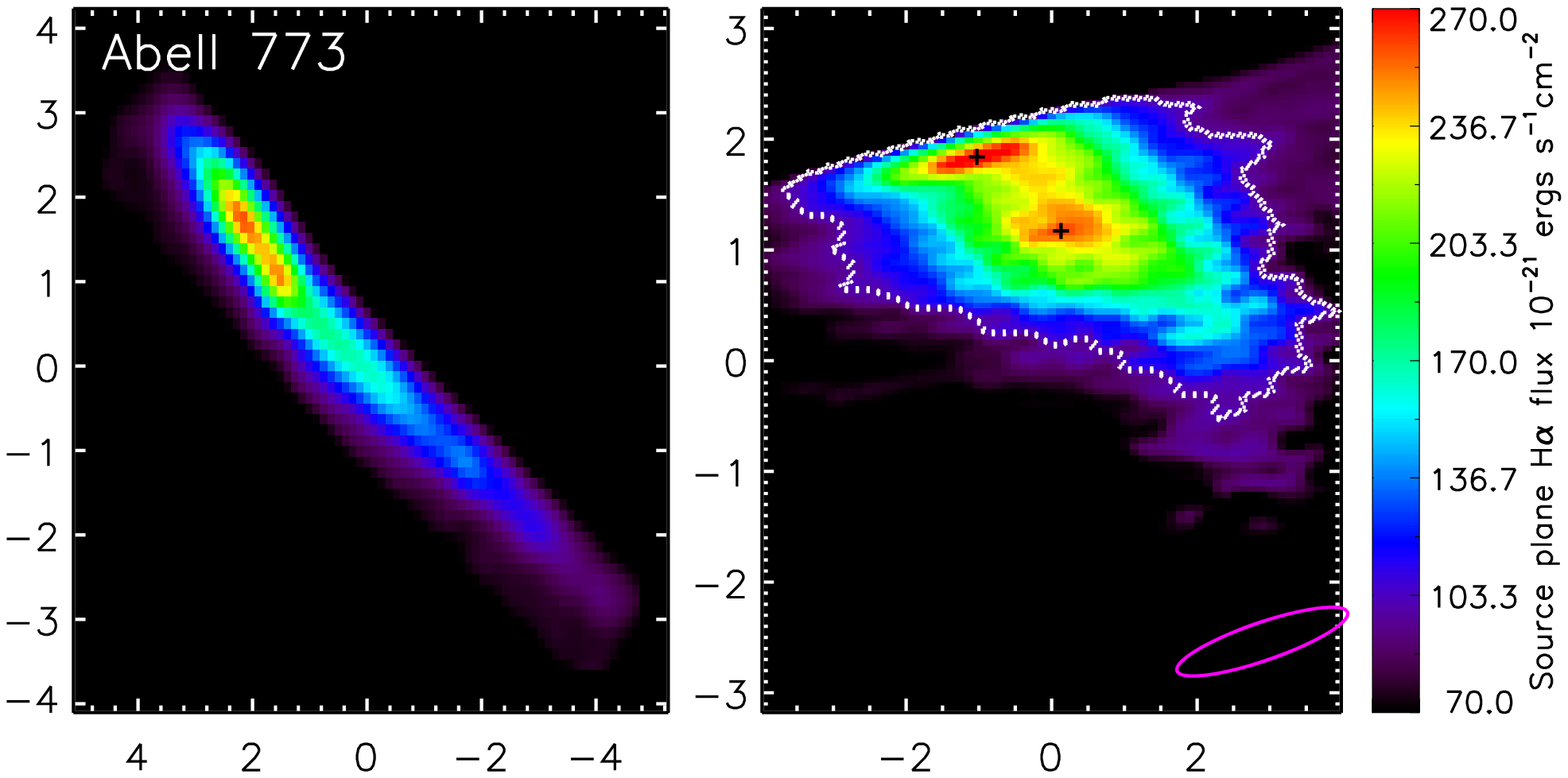}
\includegraphics[height=88mm, angle=90]{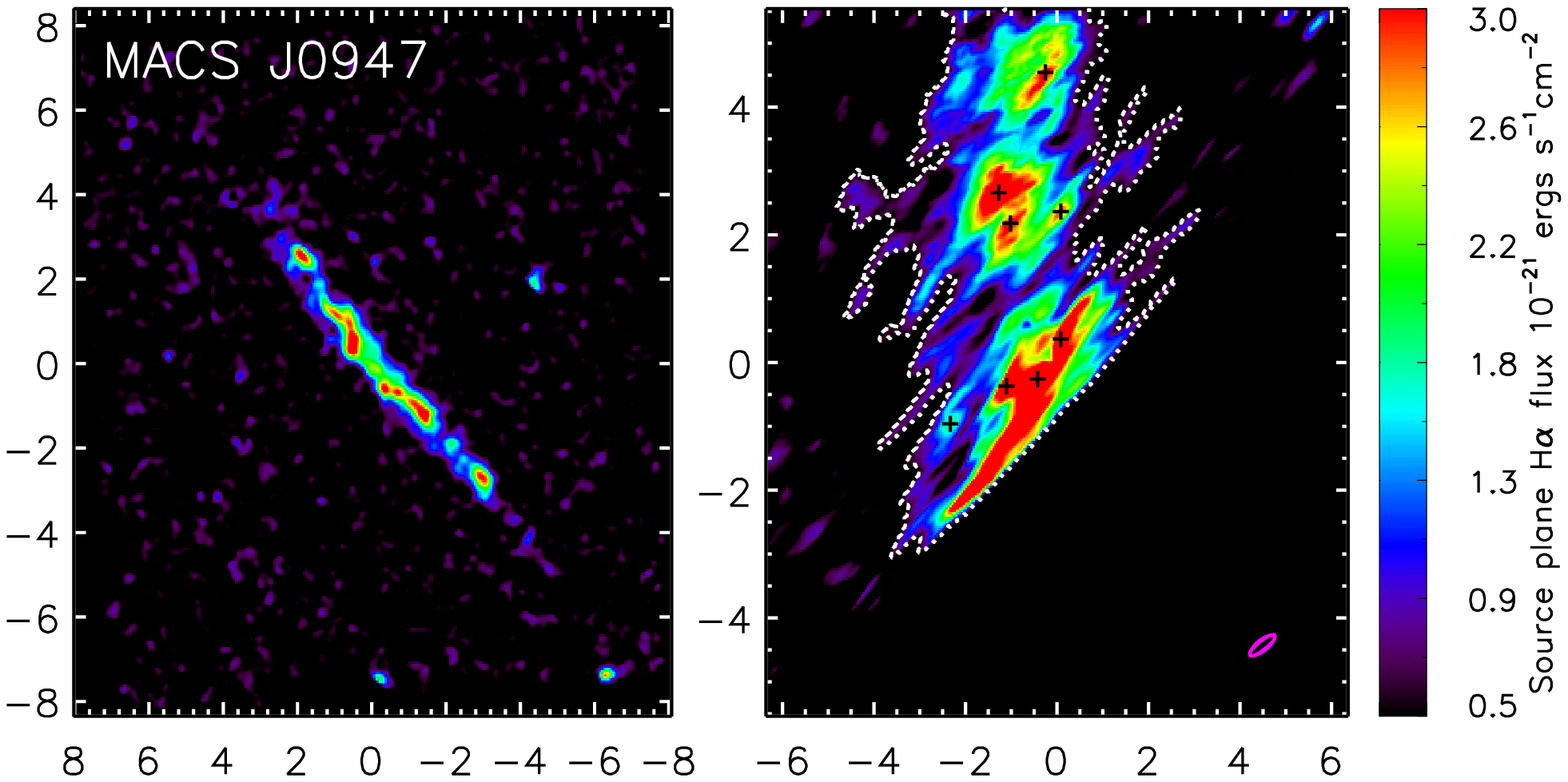}
\includegraphics[height=88mm, angle=90]{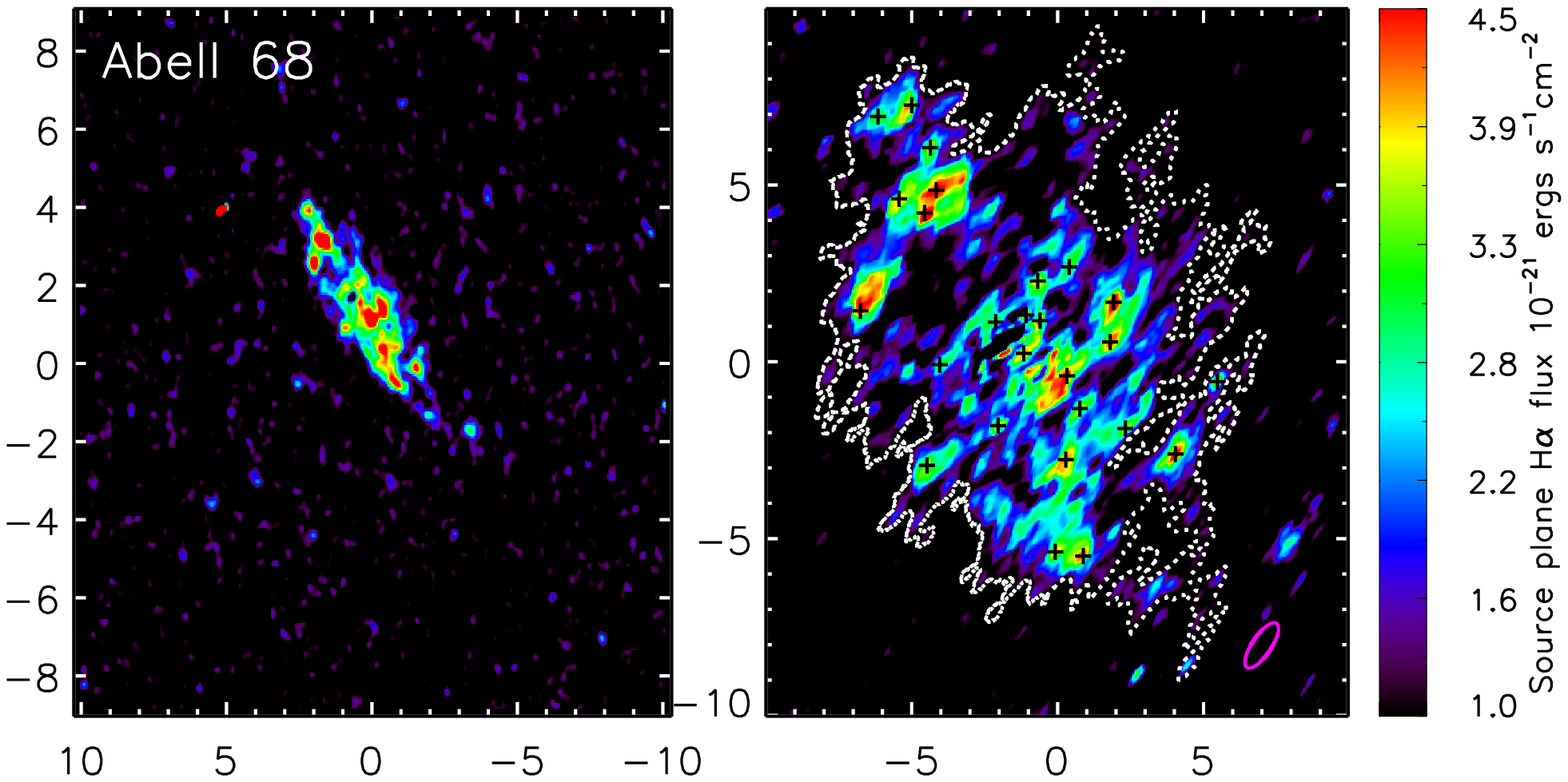}
\includegraphics[height=88mm, angle=90]{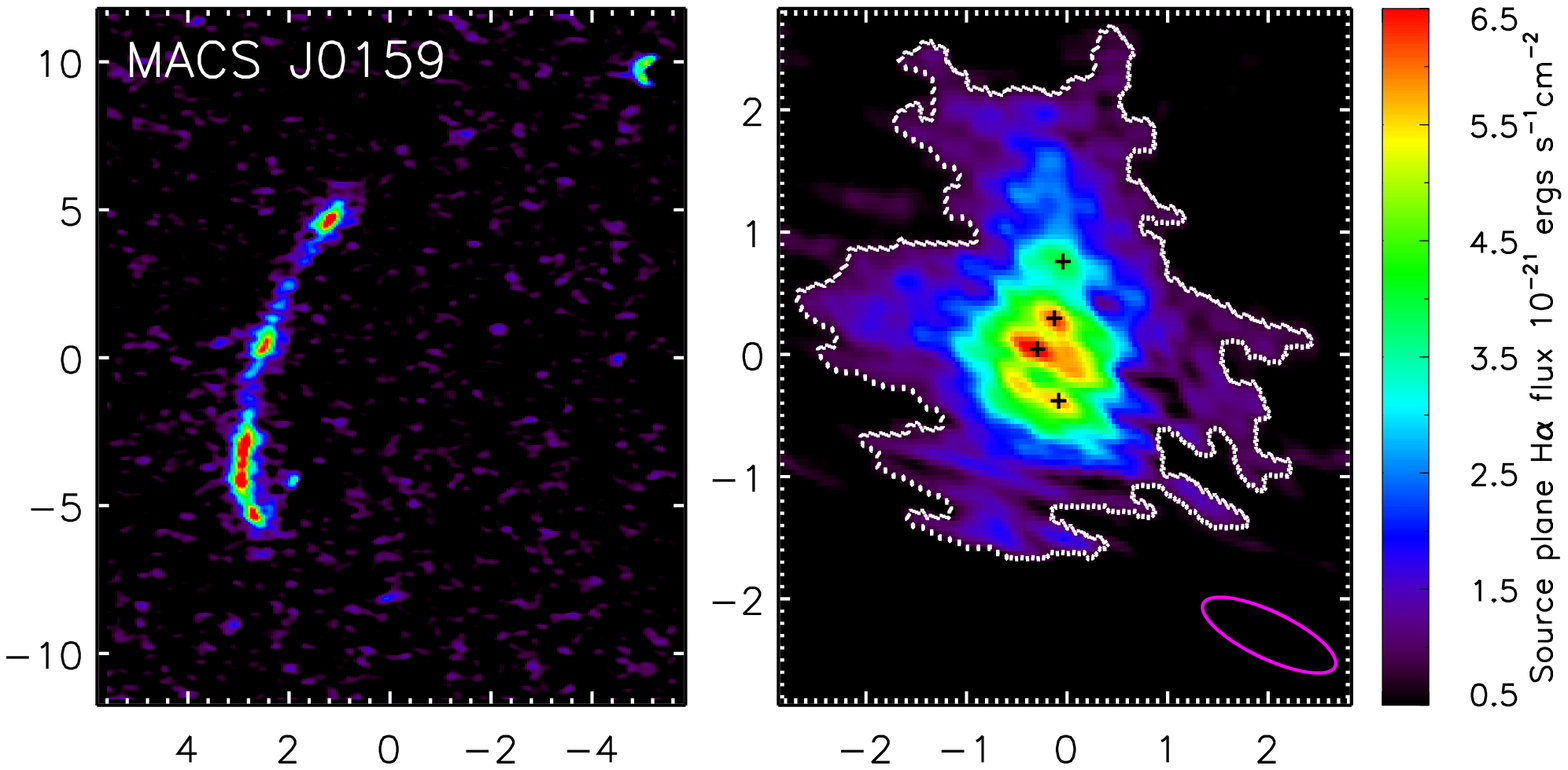}
\includegraphics[height=88mm, angle=90]{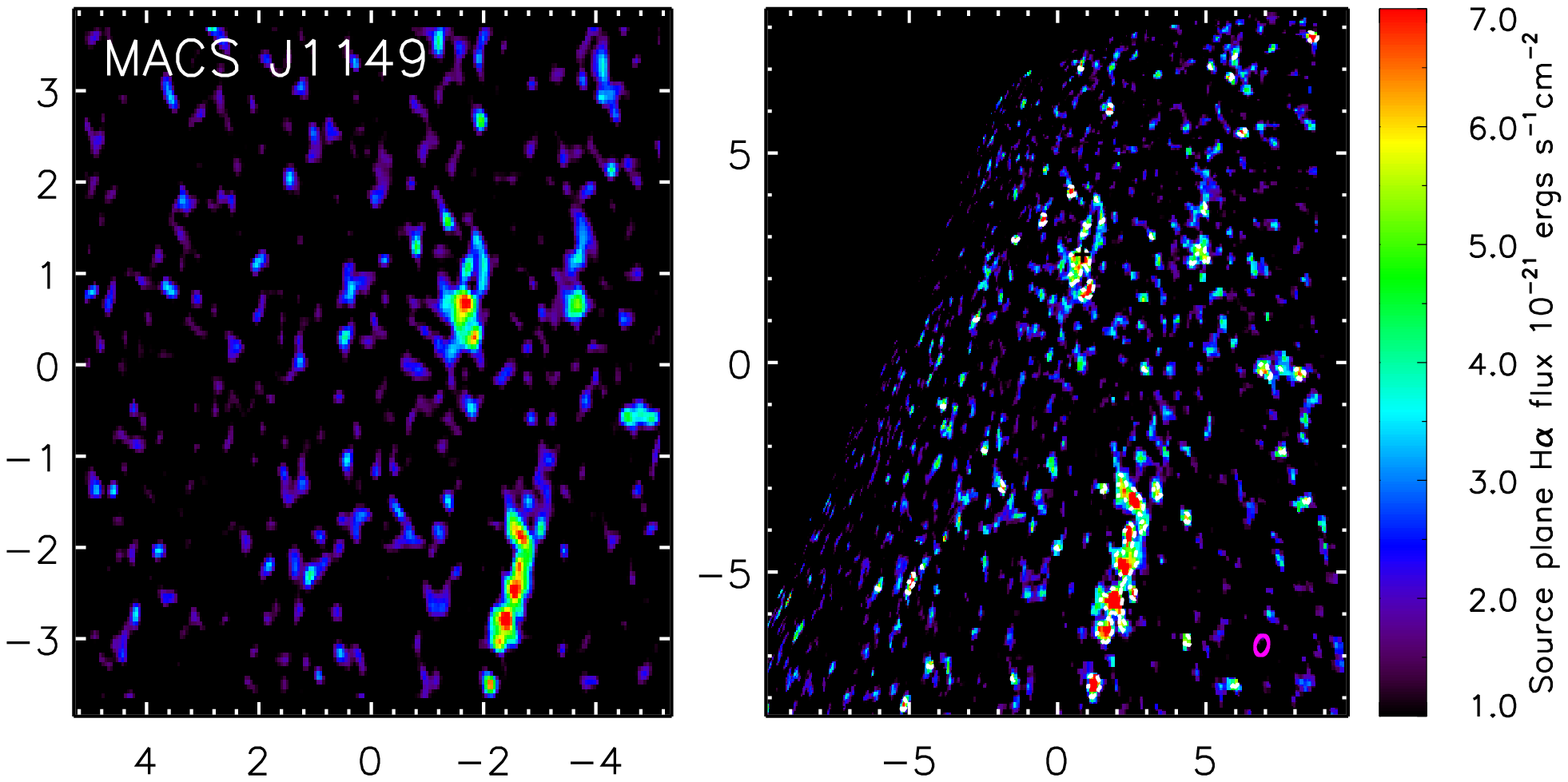}
\includegraphics[height=88mm, angle=90]{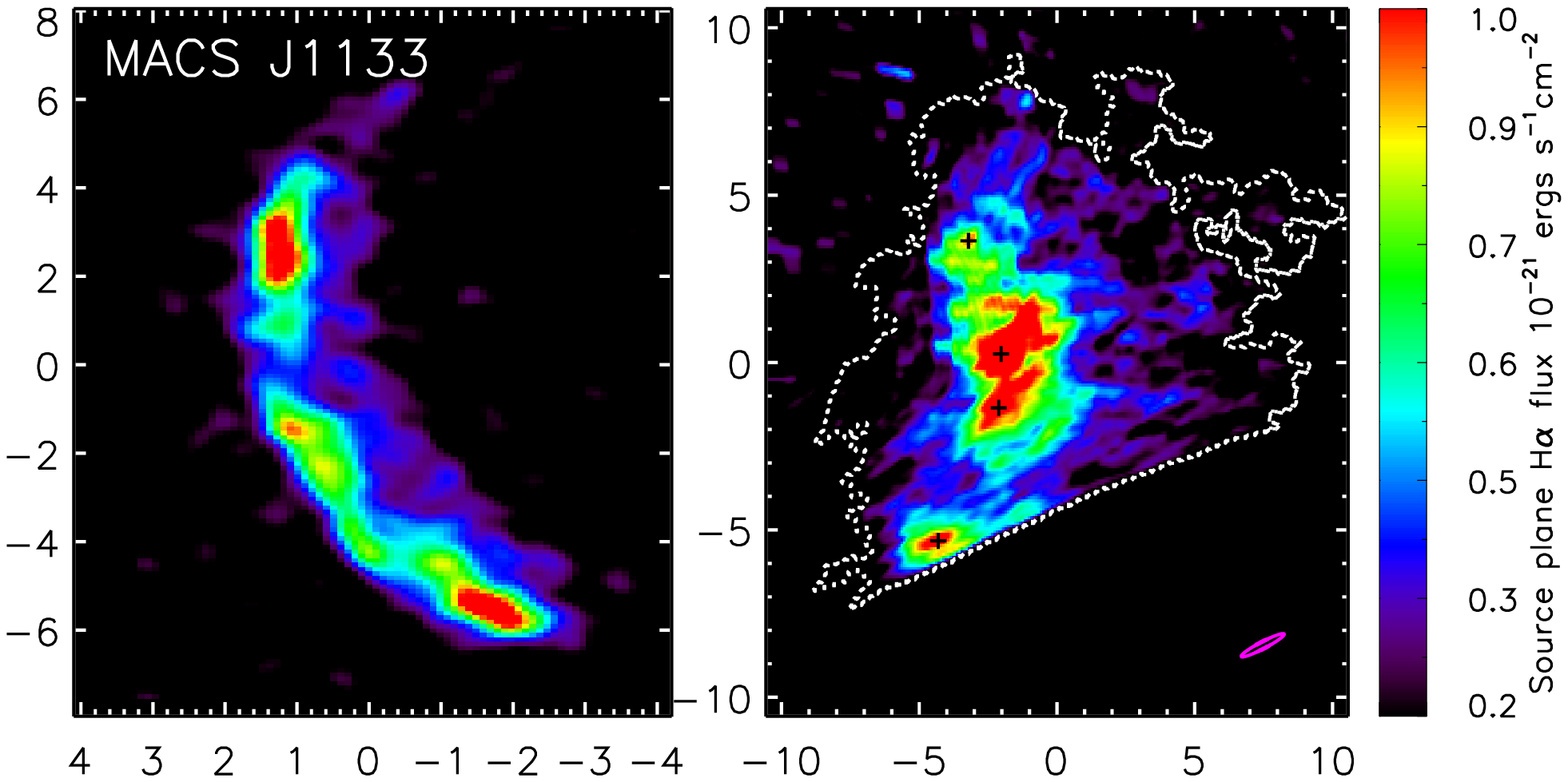}
\caption{H$\alpha$ excess images in the image plane (left) and reconstructed in the source plane (right). The image scales are in arcseconds in the image plane and in kpc in the source plane. Identified clumps are indicated in the source-plane images by black crosses, and the magenta ellipse shows the FWHM of the effective source-plane PSF, as described in the text.}
\label{fig:Ha}
\end{figure*}

\begin{sidewaystable}\centering
  \begin{tabular}{l c c c c c c c c c l}
    \hline
    Target cluster& \multicolumn{2}{c}{Arc position} & $z$ & H$\alpha$ flux & \multicolumn{2}{c}{Magnification} & Resolution & Broadband & Narrowband & Lens\\
     & RA & Dec & & (intrinsic) & $\mu_x \times \mu_y \left(\rmn{PA}\right)$ & $\mu$ & (pc) & filter & filter & model\\
    & (J2000) & (J2000) & & \small{($10^{-18}$erg/s/cm$^{2}$)} & & & & & & reference\\
    \hline
    Abell 611 & 08:00:57.30 & +36:03:37.0 & 0.908 & $30 \pm 5$ & $10.4\times 2.7\, (1\degr)$ & $28 \pm 5$ & 338 & F125W$^a$ & F126N & [1] \\
    Abell 2390 & 21:53:34.55 & +17:42:02.4 & 0.912 & $39 \pm 6$ & $5.5\times 2.3\, (73\degr)$ & $12.6 \pm 1.9$ & 435 & F125W$^b$ & F126N$^b$ & [2] \\
    Abell 773 & 09:17:58.80 & +51:43:42.3 & 1.010 & $274 \pm 49$ & $7.0\times 1.0\, (61\degr)$ & $7 \pm 1$ & 336 & F110W & F132N & [1]\\
    &&&&&&&& \& F160W$^c$ && \\
    MACS J0947.2+7623 & 09:47:15.26 & +76:23:02.9 & 1.012$^d$ & $8.5 \pm 2.2$ & $3.0\times17.7\, (51\degr)$ & $53 \pm 14$ & 172 & F125W & F132N & [3]\\
    Abell 68 & 00:37:04.91 & +09:10:21.0 & 1.017 & $119 \pm 11$ & $3.0\times1.7\, (41\degr)$ & $5.1 \pm 0.5$ & 615 & F110W & F132N & [4] \\
    &&&&&&&& \& F160W$^e$ && \\
    MACS J0159.8-0849 & 01:59:04.68 & -34:13:03.4 & 1.488$^e$ & $7 \pm 1$ & $10.7\times 3.0\, (111\degr)$ & $32 \pm 4$ & 592 & F160W & F164N & [3]\\
    MACS J1149.5+2223 & 11:49:35.30 & +22:23:45.8 & 1.490$^f$ & $11 \pm 2$ & $4.5\times 3.5\, (140\degr)$ & $15 \pm 3$ & 315 & F160W$^b$ & F164N & [5]\\
    MACS J1133.2+5008 & 11:33:14.31 & +50:08:39.7 & 1.550 & $8 \pm 1$ & $1.1\times12.7\,(67\degr)$ & $14 \pm 2$ & 68 & F160W & F167N & [6]\\
    \hline
\end{tabular}
\caption{ Properties of the redshift-selected sample. Lensing magnifies the image by a factor $\mu_x$ at a position angle PA, with a transverse magnification $\mu_y$. The total magnification factor $\mu$ is calculated from the amplification of H$\alpha$ flux, and the resolution given is the highest achievable along the most magnified direction, calculated from the FWHM of a star under the same lensing transformation as that applied to the galaxy, as described in the text. All observations were obtained under Program 12197 (Cycle 18, PI:Richard) unless otherwise stated. \newline $^a$\,Program 12065-9  $^b$\,Program 11678  $^c$\,Program 11591  $^d$\,\protect \citet{2010MNRAS.407...83E}  $^e$\,Ebeling et~al. in prep  $^f$\,\protect \citet{2007ApJ...661L..33E}.\newline [1] \protect \mbox{\citet{2010MNRAS.404..325R}}, [2] \protect \mbox{\citet{1991ApJ...366..405P}}, [3] Richard et~al. in prep, [4] \protect \mbox{\citet{2007ApJ...662..781R}}, [5] \mbox{\citet{2009ApJ...707L.163S}}, [6] \protect \mbox{\citet{2005ApJ...627...32S}}}
\label{tab:props}
\end{sidewaystable}

Colour \emph{HST} images of the clusters are shown in Figure
\ref{fig:hst}, with the critical lines at the redshift of the target
arc overlaid. We use the transformation between image and source plane
mapping from the best-fit cluster mass models (for details of the
mass models, see references in Table {\ref{tab:props}}) with {\sc
lenstool} \citep{1993PhDT.......189K,2007NJPh....9..447J} to
reconstruct the images in the source plane, and show these in Figure
\ref{fig:Ha}. In order to reconstruct the source plane morphology, {\sc lenstool} uses the mapping between the image and source planes on a cluster-by-cluster basis and ray-traces the galaxy image. The lensing effect is to stretch the galaxy image - in most cases along one direction - and so the reconstruction cannot `create' new H{\sc ii} regions, but rather the lensing has acted to extend them. As surface brightness is conserved by lensing, we then apply this conservation to obtain the intrinsic source plane flux. The total magnification is then simply the ratio of the image- to source-plane flux. To obtain the errors on the magnification, we use the family of best fit lens models which adequately describe the cluster potential, derived by sampling the posterior probability
distribution of each parameter of the model (see
\mbox{{\citet{2010MNRAS.404..325R}}} for more details). For each acceptable lens model, we reconstruct the arc and remeasure the amplification. We give the resulting magnification factors, $\mu$, and associated errors in Table {\ref{tab:props}}.

In cases where
the target is multiply-imaged, the images were reconstructed separately
and then adjusted for small differences in position and orientation
before being combined. For MACS~J0159, which consists of five images,
only the first three were used due to the high magnification gradients
in the fourth and fifth images resulting in high distortion in the
source plane reconstructions. In the case of Abell~611, we use only the
northernmost arc due to high distortion by a foreground galaxy lying
close to the line of sight of the southern arc.

\begin{figure}
\includegraphics[height=84mm, angle=90]{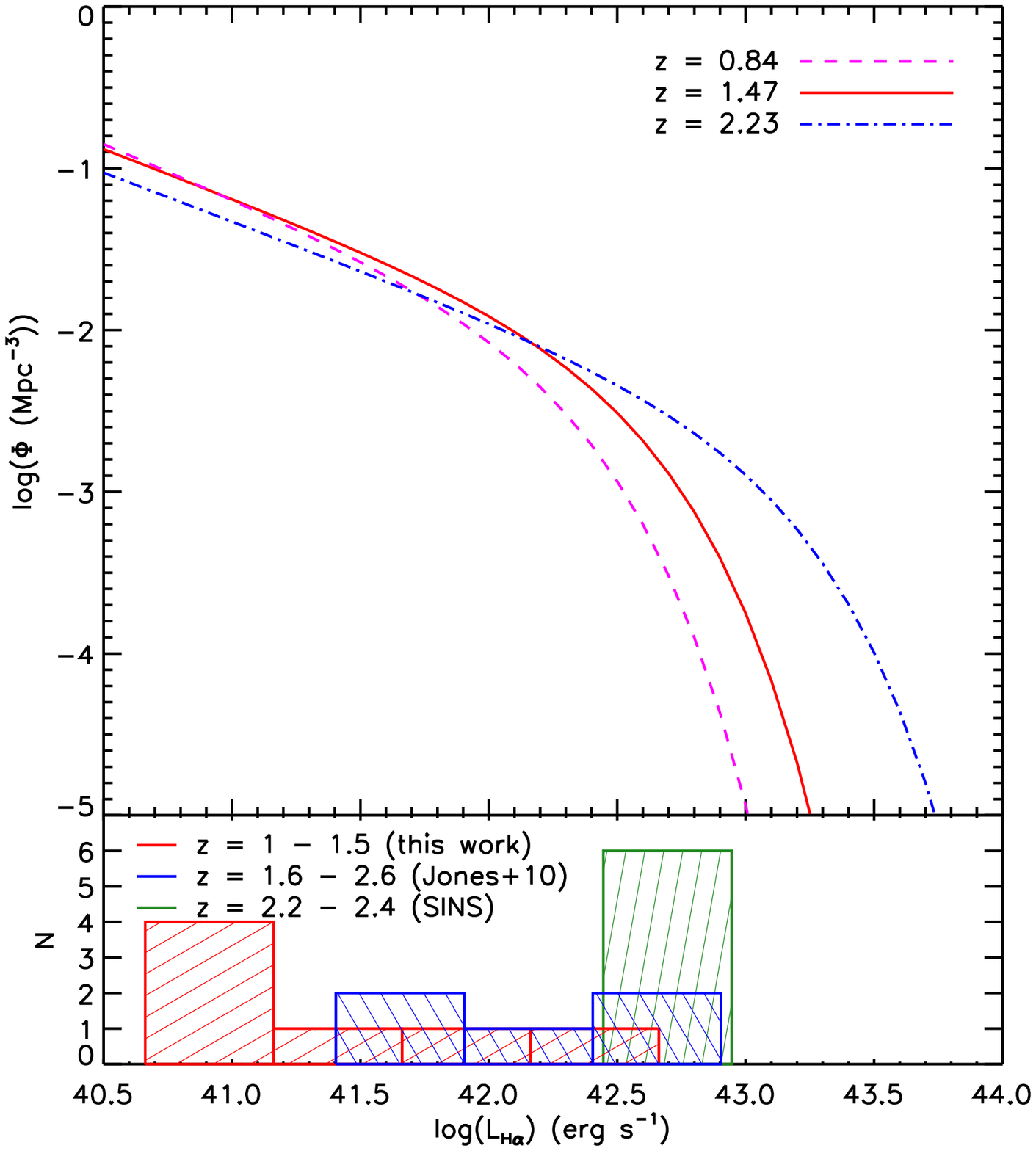}
\caption{Intrinsic H$\alpha$ luminosities of the high-$z$ samples compared to H$\alpha$ luminosity functions from HiZELS \protect \citep{2011arXiv1109.1830S}. Also shown is the range of H$\alpha$ luminosities of the \protect \citet{2011ApJ...739...45F} sample from the SINS survey at $z \sim 2$. The two lensed samples overlap in luminosity and are both at the faint end of the luminosity function, with the median of the $z \sim 1 - 1.5$ WFC3 sample lower than that of the $z = 1.6-2.6$ OSIRIS sample by a factor of 6.6$\times$, while the unlensed SINS galaxies cover a range of higher H$\alpha$ luminosities.}
\label{fig:lumfunc}
\end{figure}

We derive total magnification factors by comparing the total luminosities of the image- and source-plane H$\alpha$ excess images. The intrinsic H$\alpha$ luminosities are in the range $0.45 - 15 \times 10^{41}$erg\,s$^{-1}$ corresponding to SFRs of $0.4 - 12$M$_{\sun}$yr$^{-1}$. These are at the faint end of the H$\alpha$ luminosity function for this redshift range (see Figure \ref{fig:lumfunc}), and probe fainter galaxies than the $z \sim 2$ sample of \citet{2010MNRAS.404.1247J}, which covers the range $2.5 - 32\times 10^{41}$erg\,s$^{-1}$, although the two samples overlap in luminosity. Due to the increased sensitivity provided by the lensing magnification, both of the lensed samples cover a lower range of intrinsic H$\alpha$ luminosities than the sample of SINS galaxies studied by \citet{2011ApJ...739...45F}, which were selected to have bright H$\alpha$ and lie in the range $28 - 43 \times 10^{41}$erg\,s$^{-1}$, making them rare, intensely star-forming galaxies. Thus, by harnessing gravitational lensing we are able to probe the more `normal' star-forming population.

Since gravitational lensing can preferentially shear one direction, we estimate the effective source-plane resolution by reconstructing the image of a star from the field repositioned to lie at the centre of the target. The maximum linear resolution, derived from the FWHM of the reconstructed star in the direction of greatest magnification, is 68--615pc with a median of 360pc, sufficient to resolve giant H{\sc ii} regions.

\subsection{Comparison samples}
\label{sec:samples}

In order to interpret our high-$z$ data, we exploit the H$\alpha$ narrowband imaging from the \emph{Spitzer} Infrared Nearby Galaxies Survey \citetext{SINGS, \citealp{2003PASP..115..928K}}, which comprises H$\alpha$ imaging of 75 galaxies with corrected SFRs of up to $11$\,M$_{\sun}$yr$^{-1}$. We use the publicly available continuum-subtracted H$\alpha$ narrowband imaging and restrict the sample to those with H$\alpha$ detections with signal-to-noise of $>5$ that have no significant defects in the galaxy images (determined by visual inspection). This restricts the SINGS sample to 41 galaxies with SFR $> 4\times 10^{-4}$\,M$_{\sun}$yr$^{-1}$.

To ensure a fair comparison, we rebin the SINGS images so that the resolution is comparable to the high-$z$ data and then threshold to the median surface brightness limit of the $z \sim 1 - 1.5$ observations. It is worth noting that thresholding the images in this manner excludes $10- 50\%$ of the total star formation. This should not affect the comparison between samples which have the same surface brightness-limit, but may serve as an indication of the fraction of star formation missed in high-$z$ observations.

To provide a comparison to local galaxies which are more actively star-forming, we use the {\sc VIMOS} H$\alpha$ imaging spectroscopy of \citet{2011A&A...527A..60R}, which includes 38 LIRGs and ULIRGs at $z < 0.13$ with spatial resolution of $130\,\rmn{pc} - 1.2\,\rmn{kpc}$ and SFR $\la 25$\,M$_{\sun}$yr$^{-1}$.

We also compare the $z \sim 1 - 1.5$ sample to the $z \sim 2$ lensed arcs of \citet{2010MNRAS.404.1247J}, which were observed with Keck/{\sc OSIRIS}. In order to provide a fair comparison, we have constructed narrowband images by summing the {\sc OSIRIS} cubes over 100\,\AA\ either side of the redshifted H$\alpha$ emission line, matching the width of the WFC3 narrowband filters. The resulting images are then corrected for lensing using the same image-to-source plane mapping as \citet{2010MNRAS.404.1247J} in order to obtain the intrinsic galaxy properties.

\subsection{Determination of galaxy properties}

The total H$\alpha$ luminosities of the galaxies in all samples are determined by summing all pixels in sky-subtracted images with signal-to-noise of $>3$. In the case of the SINGS galaxies, each image was checked by visual inspection and any foreground sources and defects masked. The resulting luminosities were then compared to the published values and found to agree to within $\sim 20\%$.

We convert H$\alpha$ luminosity to SFR using the \citet{1998ARA&A..36..189K} prescription, corrected to a \citet{2003PASP..115..763C} IMF, which reduces the SFR by a factor of 1.7$\times$. As we do not have constraints on the dust extinction, we adopt an estimate of $A_{\rmn{H}\alpha}=1$ in all samples. This assumption is widely used in the literature although it is the subject of some disagreement. \mbox{\citet{2010MNRAS.402.2017G}} suggest a luminosity-dependent $A_{\rmn{H}\alpha}$ is more appropriate; were we to adopt their relation, we would obtain $A_{\rmn{H}\alpha}=0.7 - 1.6$ with a median $A_{\rmn{H}\alpha}=1.15$. However, we also note that recent work by \mbox{\citet{2012arXiv1206.1867D}} suggests that galaxies with $L_{\rmn{H}\alpha} \la 4 \times 10^{41}$\,erg\,s$^{-1}$ may be consistent with having $A_{\rmn{H}\alpha}=0$, and that above this threshold extinction increases in a luminosity-dependent way. Had we adopted this correction instead, the SFRs of the majority of our galaxies would be reduced by a factor of 2.5$\times$. The exceptions are the three brightest $z \sim 2$ galaxies, in which the SFRs would increase by factors of $1.3 - 1.8\times$, and the $z \sim 1$ galaxies Abell~68 and Abell~773; the former would be a factor of $1.8\times$ lower, while the latter would be unchanged. Qualitatively, there is no significant impact on our results, as adopting either luminosity-dependent extinction relation would serve to increase the evolution we observe in \S {\ref{sec:clumpprops}}. For simplicity and reproducability, we adopt $A_{\rmn{H}\alpha}=1$ throughout.

We define the sizes of the galaxies as twice the half-light radius. The half-light radius is determined using the continuum images to find the shape (i.e. the centre and major to minor axis ratio of an ellipse that best fits the galaxy), and then adjusting the semi-major axis of the ellipse until it encompasses half of the total H$\alpha$ luminosity calculated in the manner described above. The galaxy-averaged star formation surface density, $\Sigma_{\rmn{SFR}}$ is defined from the total luminosity enclosed within two half-light radii per unit area.

\section{Results and Analysis}
\label{sec:results}

\subsection{The Spatial Distribution of Star Formation}
A common theme in the recent literature is that high redshift galaxies
are ``clumpier'' than galaxies in the local universe. This concept
originates from the frequent appearance of ``chain'' galaxies in the
high redshift universe \citep[e.g.][]{1995AJ....110.1576C,2004ApJ...612..191E,2005ApJ...627..632E}. Even without looking at the properties of
individual star forming regions, it is interesting to compare the
morphologies of the star-forming regions across the samples.

\begin{figure*}
\includegraphics[width=168mm, angle=0]{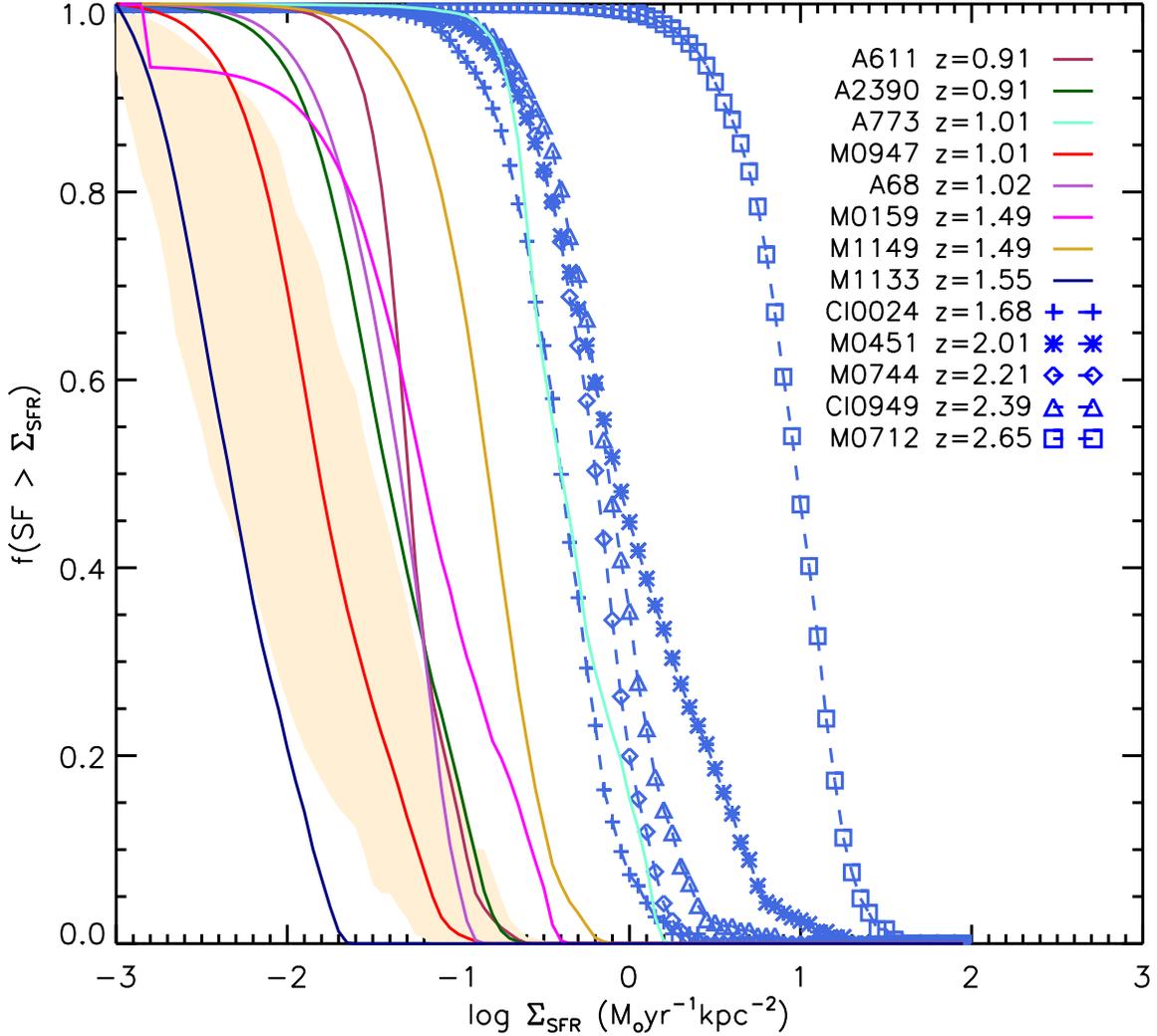}
\caption{The fraction of star formation within each galaxy occurring above a given surface brightness, for the $z \sim 1 - 1.5$ and $z \sim 2$ samples. The shaded region is the interquartile range of the SINGS $z \sim 0$ sample. There are two galaxies, MACS~J1133 and MACS~J0947, from the $z \sim 1 - 1.5$ sample with similar surface brightnesses to the $z=0$ sample, and the remainder are systematically brighter. The $z \sim 2$ sample has significantly higher surface brightnesses. Hence, there is clear evolution in the surface brightnesses of galaxies with redshift.}
\label{fig:sfdenhist}
\end{figure*}

From visual inspection, it is clear that there are significant
differences between the samples. In particular, the surface brightness
distributions of the galaxies show distinct differences in the
different samples.  In Fig.~\ref{fig:sfdenhist}, we show the fraction
of star formation in pixels above a given $\Sigma_{\rmn{SFR}}$ for the
$z \sim 1 - 1.5$ and $z \sim 2$ samples, with the interquartile range of the
thresholded SINGS sample shown for comparison.

To allow for the differing surface brightness limits of the samples, we only show star formation above a surface brightness of $\Sigma_{\rmn{SFR}} = 0.001$\,M$_{\sun}$\,yr$^{-1}$\,kpc$^{-2}$. This enables us to compare the star formation occurring in bright regions in a consistent manner. From the peaks - i.e. the points at which the lines tend to zero - we can see that the samples are different, with the $z \sim 2$ galaxies having peak surface brightnesses of around an order of magnitude higher than the lower-$z$ samples. Similarly, the $z \sim 1 - 1.5$ sample is systematically brighter than SINGS sample, with the exception of MACS~J1133, which is similar to the fainter $z=0$ galaxies, MACS~J0947 which is similar to the median of the $z=0$ sample, and Abell~773 which appears similar to the $z \sim 2$ galaxies.

As a statistical measure of the clumpiness of galaxies, we investigate
using the Gini coefficient, $G$, which is used in economics to measure
the inequality of wealth in a population \citep{gini-1912}. It has
values from 0 to 1, where at the extremes $G = 0$ for a completely
uniform distribution, and $G=1$ if there is only one non-zero
value. Following \citet{2011ApJ...731...65F}, we use it to quantify the
distribution of flux in an image, so a value close to one indicates
that the profile has a single peak (in the case of $G=1$, all of the
flux would be in a single pixel), a galaxy with multiple clumps would
have a lower $G$, and at the extreme, a galaxy with completely uniform
surface brightness would have $G=0$.

In the $z \sim 1 - 1.5$ sample, we find a narrow range of $0.25 \leq G \leq
0.39$ with a median of $G = 0.34$. The $z \sim 2$ sample is marginally
higher, with $0.42 \leq G \leq 0.56$ and a median of $G = 0.43$. The
$z=0$ SINGS sample has a similar median $G = 0.45$ but a much wider
range of $0.05 \leq G \leq 0.82$, and the low-$z$ ULIRGs have $0.38
\leq G \leq 0.85$ with the highest median $G = 0.70$. On the basis of
the Gini coefficient there are no clear differences between the
samples.  Comparing the Gini coefficients with the visual appearance of
the galaxies, the lack of distinction reflects the fact that a low Gini
coefficient may arise from either a smooth distribution of star
formation or from star formation that is concentrated into a large
number of distinct clumps.  Furthermore, we find no strong correlations
between $G$ and any of the properties of the galaxies. Clearly, to
progress further we will have to compare the properties of individual
clumps. In particular, we will show that the clump luminosity function
provides a good means of distinguishing different galaxy star formation
morphologies.

\subsection{Properties of star-forming clumps}

\subsubsection{Definition of Clumps}
\label{sec:clumpdef}

Studies of H{\sc ii} regions or star-forming clumps have used a variety of methods to define and separate clumps from the background emission of the galaxy. Usually an isophote is defined at 3$\sigma$ above the background noise \citep[e.g.][]{1997ApJS..108..199G,2010MNRAS.404.1247J}. However, this method is clearly dependent on the noise properties of the image, and thus is problematic when comparing local and high-redshift observations. In particular, as high-redshift galaxy images tend to have high relative noise levels and low dynamic range, the choice of isophote tends to select only the brightest regions in the galaxy, neglecting any lower-surface brightness clumps and underestimating their sizes.

An alternative is the IRAF task \verb'daofind' as employed by \citet{2011ApJ...739...45F}, which is designed to locate point sources in images. However, we found that it did not perform well on our sample. This is likely to be because \verb'daofind' requires an expected size of features to look for. As the clumps of \citet{2011ApJ...739...45F} are largely unresolved, they were able to use the PSF of their observations as the expected size. As our clumps are resolved, the routine does not work reliably. For this paper, we therefore use the 2D version of \verb'clumpfind' \citep{1994ApJ...428..693W}, which uses multiple isophotes to define clumps. We defined the contour levels with respect to the rms noise in the image, starting at 3$\sigma$ and increasing in 1$\sigma$ intervals until the peak value of the image is reached. The data are first contoured at the highest level to locate clumps, and the algorithm then works down in brightness through the contour levels. Any isolated contours are defined as new clumps, while others extend existing clumps. If a contour surrounds one existing peak, they are allocated to that clump, and any which enclose two or more are divided using a `friends-of-friends' algorithm. The advantages of this approach are that it enables a consistent clump definition to be applied to multiple data sets, lower surface brightness clumps are not excluded, and there is no assumption made about the clump profile.

The clumps identified by \verb'clumpfind' were all confirmed by visual inspection to remove any sources not associated with the target galaxy, of particular importance in the case of the SINGS images where foreground sources lie close to or overlap the target galaxies. The area $A$ of the clump is then obtained from the number of pixels assigned to it, multiplied by the source-plane pixel scale, and from this we define the effective radius $r = \sqrt{A/\pi}$. We only accept clumps where $2r$ is larger than the FWHM of the PSF, so all clumps are resolved. 

\begin{figure}
\includegraphics[height=84mm, angle=90]{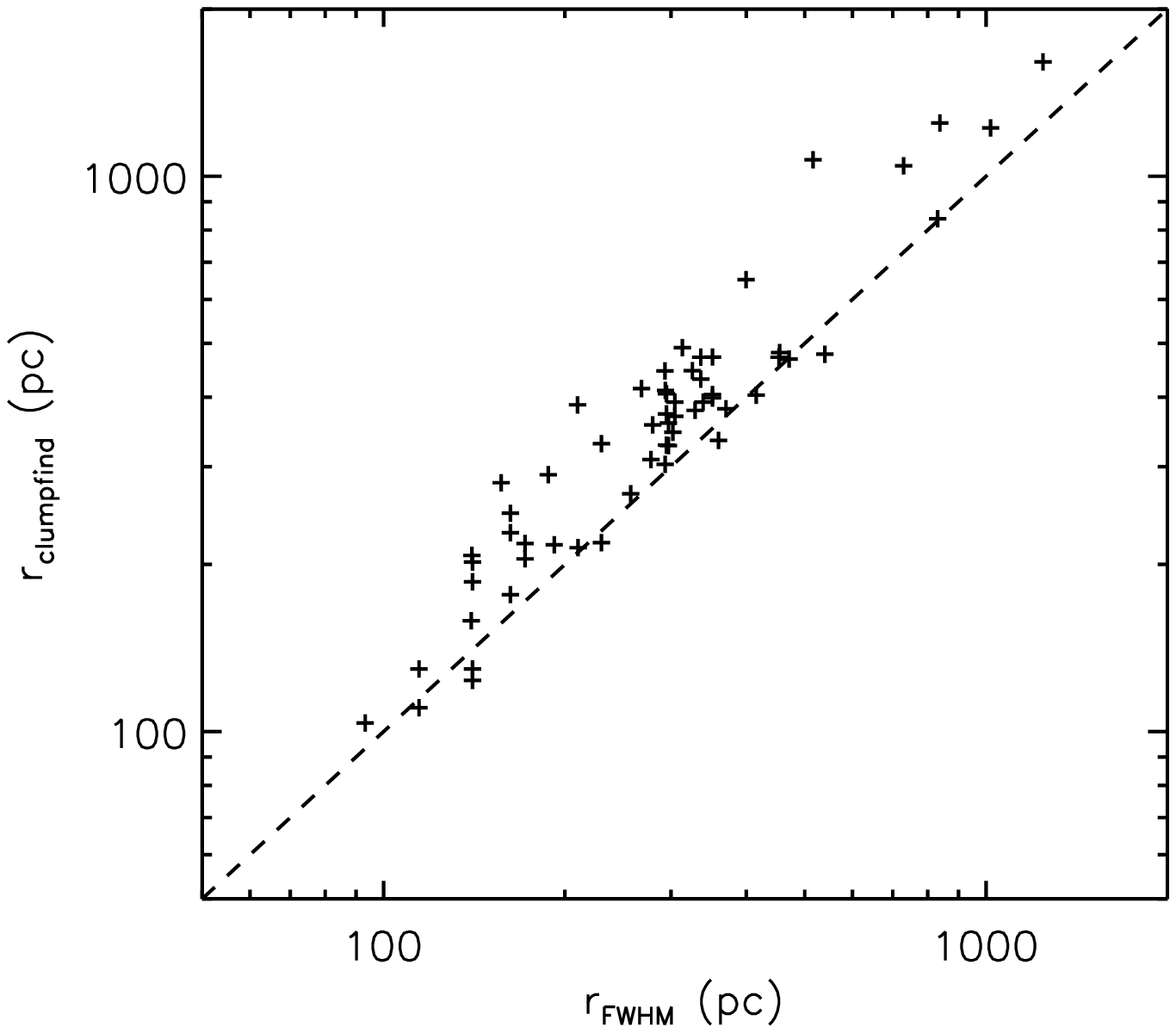}
\caption{Comparison of the clump size $r_{\rmn{clumpfind}}$ output by clumpfind with the size $r_{\rmn{FWHM}}$ obtained by taking the FWHM of a 2D Gaussian profile fit. On average, clumpfind outputs sizes 25\% larger than the FWHM. For consistency, we adopt the isophotal size output by $r_{\rmn{clumpfind}}$ in all samples.}
\label{fig:rcomp}
\end{figure}

Due to the manner in which clumps are `grown,' their sizes returned by \verb'clumpfind' tend to be larger than those obtained by other methods. As a comparison, we also fit a 2D elliptical Gaussian profile to each peak and measure the FWHM. A comparison of the clump radii found by the two methods is shown in Figure \ref{fig:rcomp}. The rms difference between the two radii is $\sim 100$pc, and on average we find that \verb'clumpfind' outputs sizes 25\% higher than the FWHM. \citet{2012MNRAS.tmp.2831W} note that sizes defined through isophotes can be unreliable due to the level of `tuning' required to select an appropriate isophote level in a given galaxy. This is less significant with \verb'clumpfind' because this tuning is not required; the use of multiple isophote levels in all galaxies allows the levels to be defined in a consistent way across a large sample. We therefore find much lower scatter between the isophotal sizes output by \verb'clumpfind' and the clump FWHM than they do in their sample. Throughout this work, we use the \verb'clumpfind' size for all samples, and give error bars that encompass the FWHM of the clumps.

The sizes and H$\alpha$-derived SFRs of the $z \sim 1 - 1.5$ clumps are given in Table \ref{tab:clumps}. We analyse these properties in comparison to the other samples below.

\begin{table}
  \caption{Properties of clumps identified in the $z \sim 1 - 1.5$ sample, determined as described in \S \ref{sec:clumpdef}.}
  \label{tab:clumps}
  \begin{tabular}{l c c}
    \hline
    Clump & radius (pc) & SFR (M$_{\sun}$\,yr$^{-1}$)\\
    \hline
MACS J0947-1 & 350 $\pm$ 56 & 0.054 $\pm$ 0.010 \\
MACS J0947-2 & 324 $\pm$ 22 & 0.0328 $\pm$ 0.0086 \\
MACS J0947-3 & 384 $\pm$ 48 & 0.045 $\pm$ 0.012 \\
MACS J0947-4 & 334 $\pm$ 39 & 0.0350 $\pm$ 0.0092 \\
MACS J0947-5 & 318 $\pm$ 38 & 0.0339 $\pm$ 0.0089 \\
MACS J0947-6 & 376 $\pm$ 6 & 0.0347 $\pm$ 0.0091 \\
MACS J0947-7 & 311 $\pm$ 17 & 0.0261 $\pm$ 0.0068 \\
MACS J0947-8 & 149 $\pm$ 9 & 0.0050 $\pm$ 0.0013 \\
MACS J0159-1 & 402 $\pm$ 89 & 0.282 $\pm$ 0.060 \\
MACS J0159-2 & 370 $\pm$ 77 & 0.203 $\pm$ 0.043 \\
MACS J0159-3 & 530 $\pm$ 130 & 0.355 $\pm$ 0.076 \\
MACS J0159-4 & 468 $\pm$ 13 & 0.170 $\pm$ 0.036 \\
Abell 611-1 & 730 $\pm$ 180 & 0.370 $\pm$ 0.083 \\
Abell 611-2 & 560 $\pm$ 160 & 0.181 $\pm$ 0.041 \\
Abell 611-3 & 630 $\pm$ 140 & 0.200 $\pm$ 0.045 \\
Abell 611-4 & 390 $\pm$ 59 & 0.065 $\pm$ 0.015 \\
Abell 68-1 & 378 $\pm$ 26 & 0.081 $\pm$ 0.016 \\
Abell 68-2 & 132 $\pm$ 8 & 0.0075 $\pm$ 0.0015 \\
Abell 68-3 & 375 $\pm$ 24 & 0.076 $\pm$ 0.015 \\
Abell 68-4 & 354 $\pm$ 25 & 0.070 $\pm$ 0.014 \\
Abell 68-5 & 509 $\pm$ 31 & 0.114 $\pm$ 0.023 \\
Abell 68-6 & 337 $\pm$ 33 & 0.062 $\pm$ 0.013 \\
Abell 68-7 & 386 $\pm$ 61 & 0.099 $\pm$ 0.020 \\
Abell 68-8 & 205 $\pm$ 42 & 0.0273 $\pm$ 0.0055 \\
Abell 68-9 & 348 $\pm$ 44 & 0.069 $\pm$ 0.014 \\
Abell 68-10 & 299 $\pm$ 89 & 0.066 $\pm$ 0.013 \\
Abell 68-11 & 328 $\pm$ 31 & 0.057 $\pm$ 0.012 \\
Abell 68-12 & 312 $\pm$ 15 & 0.0476 $\pm$ 0.0097 \\
Abell 68-13 & 412 $\pm$ 60 & 0.100 $\pm$ 0.020 \\
Abell 68-14 & 293 $\pm$ 16 & 0.0429 $\pm$ 0.0087 \\
Abell 68-15 & 263 $\pm$ 5 & 0.0302 $\pm$ 0.0061 \\
Abell 68-16 & 280 $\pm$ 50 & 0.0454 $\pm$ 0.0092 \\
Abell 68-17 & 169 $\pm$ 7 & 0.0132 $\pm$ 0.0027 \\
Abell 68-18 & 195 $\pm$ 33 & 0.0215 $\pm$ 0.0044 \\
Abell 68-19 & 224 $\pm$ 6 & 0.0199 $\pm$ 0.0040 \\
Abell 68-20 & 239 $\pm$ 51 & 0.0344 $\pm$ 0.0070 \\
Abell 68-21 & 135 $\pm$ 5 & 0.0069 $\pm$ 0.0014 \\
Abell 68-22 & 163 $\pm$ 23 & 0.0142 $\pm$ 0.0029 \\
Abell 68-23 & 112 $\pm$ 2 & 0.00489 $\pm$ 0.00099 \\
Abell 68-24 & 171 $\pm$ 31 & 0.0161 $\pm$ 0.0033 \\
Abell 68-25 & 98 $\pm$ 5 & 0.00416 $\pm$ 0.00085 \\
Abell 68-26 & 122 $\pm$ 8 & 0.0066 $\pm$ 0.0013 \\
Abell 2390-1 & 352 $\pm$ 59 & 0.115 $\pm$ 0.025 \\
Abell 2390-2 & 404 $\pm$ 68 & 0.136 $\pm$ 0.030 \\
Abell 2390-3 & 409 $\pm$ 6 & 0.086 $\pm$ 0.019 \\
Abell 2390-4 & 366 $\pm$ 26 & 0.067 $\pm$ 0.015 \\
Abell 2390-5 & 463 $\pm$ 9 & 0.111 $\pm$ 0.024 \\
Abell 2390-6 & 347 $\pm$ 13 & 0.059 $\pm$ 0.013 \\
Abell 2390-7 & 470 $\pm$ 1 & 0.093 $\pm$ 0.020 \\
Abell 2390-8 & 341 $\pm$ 73 & 0.069 $\pm$ 0.015 \\
Abell 773-1 & 1040 $\pm$ 200 & 5.6 $\pm$ 1.3 \\
Abell 773-2 & 1430 $\pm$ 180 & 9.6 $\pm$ 2.2 \\
MACS J1133-1 & 1120 $\pm$ 100 & 0.118 $\pm$ 0.025 \\
MACS J1133-2 & 890 $\pm$ 160 & 0.068 $\pm$ 0.015 \\
MACS J1133-3 & 790 $\pm$ 280 & 0.057 $\pm$ 0.012 \\
MACS J1133-4 & 835 $\pm$ 4 & 0.0334 $\pm$ 0.0072 \\
MACS J1149-1 & 174 $\pm$ 34 & 0.084 $\pm$ 0.020 \\
\hline
\end{tabular}
\end{table}

\subsubsection{Clump properties}
\label{sec:clumpprops}

One way of quantifying the `clumpiness' of a galaxy is to consider the fraction of a galaxy's total H$\alpha$ luminosity contained within clumps. We find medians of 31\% in SINGS, 36\% for the $z<0.13$ ULIRGs, 50\% for the $z \sim 1 - 1.5$ sample and 68\% for the $z \sim 2$ sample. Thus, as expected, the higher-$z$ galaxies are clumpier than their local counterparts.

\begin{figure*}
\includegraphics[height=168mm, angle=90]{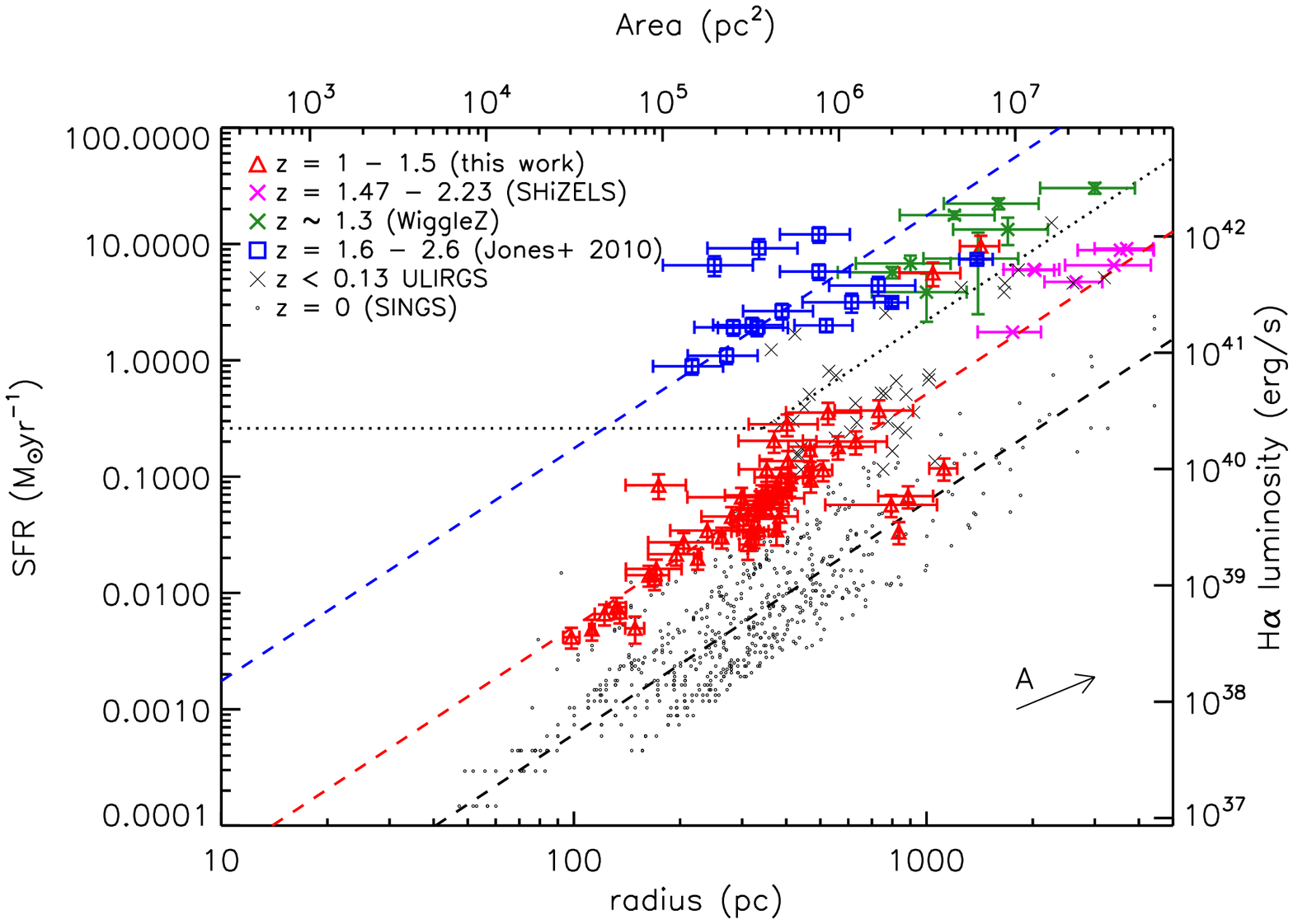}
\caption{H$\alpha$ SFR for extracted H{\sc ii} regions as a function of size, compared to the lensed $z \sim 2$ sample of \protect \citet{2010MNRAS.404.1247J}, high-$z$ unlensed samples from SHiZELS (Swinbank et al. in prep) and WiggleZ \protect \citep{2012MNRAS.tmp.2831W}, low-$z$ ULIRGs from \protect \citet{2011A&A...527A..60R} and the $z=0$ SINGS galaxies \protect \citep{2003PASP..115..928K}. Star formation rates are calculated using the \citet{1998ARA&A..36..189K} prescription adjusted for a Chabrier IMF with a dust extinction $A_{\rmn{H}\alpha} = 1$ in all samples, and the error bars of the high-$z$ lensed sources are dominated by the uncertainty in the lensing magnification. Dashed lines show the median surface brightnesses in the SINGS, $z \sim 1 - 1.5$ and $z \sim 2$ samples. The black dotted line indicates the sensitivity limit of the $z \sim 2$ OSIRIS observations. The arrow indicates the effect of degrading the image resolution, as discussed in the text. The four lowest surface brightness clumps in the $z \sim 1 - 1.5$ sample come from one galaxy (MACS~J1133), and the two brightest regions are from Abell~773, the most compact galaxy in the sample. The remaining galaxies have clumps with surface brightnesses in between those of the $z=0$ and $z \sim 2$ samples, similar to local ULIRGs.}
\label{fig:sizelum}
\end{figure*}

We now consider the properties of the clumps themselves, and first compare the H$\alpha$-derived SFR to the clump radius, as shown in Figure \ref{fig:sizelum}. Locally, there is a well-defined relationship between these properties, as found by \citet{1988ApJ...334..144K} who found almost constant surface brightness in local H{\sc ii} regions, except in merging and interacting systems \citep{2006A&A...445..471B}. The situation at high-$z$, though, appears different; \citet{2009MNRAS.400.1121S} and \citet{2010MNRAS.404.1247J} found clumps with SFRs of $\sim 100\times$ higher at a given size than found locally, in systems with no evidence of interactions.

Figure~\ref{fig:sizelum} is an updated version of one presented in \citet{2010MNRAS.404.1247J}, where we have re-analysed the $z \sim 2$ and SINGS galaxies using \verb'clumpfind' so that clumps are defined consistently across all samples, and we have added the results from our new $z \sim 1 - 1.5$ data set and the $z < 0.13$ ULIRGS as well as the $z = 1-2$ results from {\sc SHiZELS} (Swinbank et~al. in prep.) and WiggleZ \citep{2012MNRAS.tmp.2831W}. We show lines of median surface brightness in the samples, and vertical offsets from these lines represent differences in the surface density of star formation, $\Sigma_{\rmn{SFR}}$, in the clumps. We will explore the relation of these offsets to global galaxy properties in Section~\ref{sec:disc}.

We note that the clumps we identify in the SINGS galaxies are derived from images which have been degraded to comparable resolution to the high-$z$ data, and we find the effect of this is to decrease the surface brightness by a factor of $\sim 2\times$, as the size increases more than the luminosity. The points in Figure~\ref{fig:sizelum} move along the vector labelled 'A'. Defining clumps in the $z=0$ sample in this way ensures the fairest possible comparison with the high-$z$ data.

Upon re-analysis using \verb'clumpfind', we find some lower $\Sigma_{\rmn{SFR}}$ regions in the \citet{2010MNRAS.404.1247J} sample, but they all remain separated from the local relation by a factor of $\sim 100\times$. This confirms the large differences between the local and high redshift population already noted by \citet{2009MNRAS.400.1121S} and \citet{2010MNRAS.404.1247J}.

Our new $z \sim 1 - 1.5$ sample fits in between the SINGS and $z\sim 2$ samples, with the exception of the two regions from the most compact source Abell~773, which have $\Sigma _{\rmn{SFR}}$ similar to the $z \sim 2$ sample, and the four regions from MACS~1133, which are similar to $z=0$ clumps. This indicates clear evolution in clump surface brightness,  $\Sigma_{\rmn{SFR}}$, with redshift.

The surface brightness limit of the $z \sim 2$ data means that we
cannot identify the low star formation rate clumps in that sample. We
show a dotted line representing the lower limit at which we define
clumps in the $z \sim 2$ galaxies. It is likely that there are
additional clumps which lie below this limit and are undetected;
however, such clumps make only a small contribution to the total SFR,
as we shall discuss in Section~\ref{sec:lf}.

Selection effects have no impact on the lack of high surface brightness
regions in the lower redshift samples, however. The intense star-forming regions are clearly more common in high-$z$ galaxies; they are found only in extreme systems such as the Antennae locally, but exist in all five of the $z \sim 2$ galaxies and one of the eight $z \sim 1 - 1.5$ sample. 

As noted in \S~\ref{sec:sample}, the $z \sim 2$ galaxies have `normal' SFRs for their redshift, below the knee of the H$\alpha$ luminosity function. The offsets seen in the figure emphasise the importance of analysing
clumps in terms of their surface brightness.  This is even more evident
if the clumps belonging to a single galaxies are examined
separately. Rather than being distributed across the plot at random,
individual galaxies form a much tighter sequence with all the clumps
sharing a common surface brightness, particularly in the low redshift
sample. Thus the spread in clump properties in Figure~\ref{fig:sizelum} appears to be
driven by \emph{global} differences in the galaxies. 

\begin{figure*}
\includegraphics[height=168mm, angle=90]{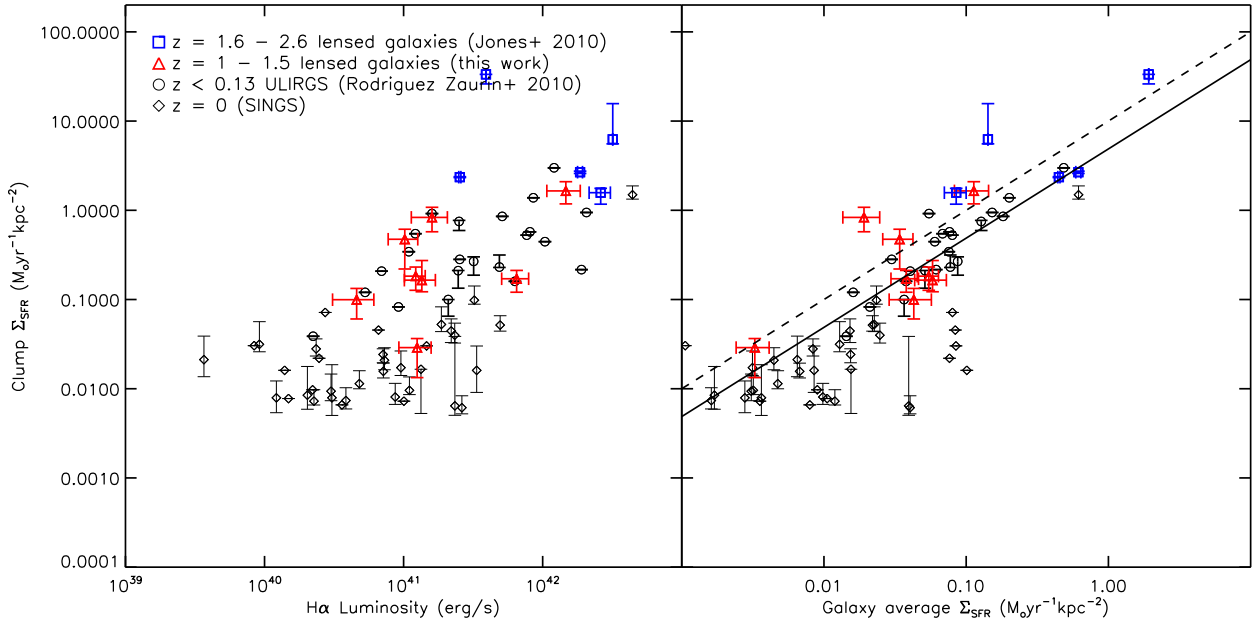}
\caption{Comparisons between the star formation surface density $\Sigma_{\rmn{SFR}}$ of star-forming clumps within each galaxy and the intrinsic H$\alpha$ luminosity and galaxy-averaged $\Sigma_{\rmn{SFR}}$. The clump $\Sigma_{\rmn{SFR}}$ shown is the median for each galaxy, with error bars encompassing the full range of $\Sigma_{\rmn{SFR}}$ for all clumps within each galaxy. The solid line is the best fit to the data, and the dashed line illustrates the clump $\Sigma_{\rmn{SFR}}$ expected from theory, discussed in Section~\ref{sec:disc}. We find that both are correlated at the 5$\sigma$ level, implying that we find more high-$\Sigma_{\rmn{SFR}}$ at high redshift because there are more high-SFR and $\Sigma_{\rmn{SFR}}$ galaxies at this epoch.}
\label{fig:clumpplots}
\end{figure*}

We therefore next compare the clump $\Sigma_{\rmn{SFR}}$ to the
properties of their host galaxies in Figure~\ref{fig:clumpplots}. In
the left-hand panel, we correlate the clump properties with the total
star formation rate of the galaxy. For clarity, we plot the median
clump $\Sigma_{\rmn{SFR}}$ in each individual galaxy, and the error
bars encompass the central 68\% of clumps within each galaxy
(i.e. 1\,$\sigma$ if they follow a Gaussian distribution) There is evidence for correlation between the clump $\Sigma_{\rmn{SFR}}$ and the
galaxy H$\alpha$ luminosity (which we assume to be proportional to the
total SFR); we find a Spearman rank correlation coefficient $\rho =
0.69$, representing a 5.8\,$\sigma$ deviation from the null hypothesis
of no correlation. This suggests that the star-formation in the
high-$z$ sample follows a similar trend to the local sample, and
that the differences seen in Fig.~\ref{fig:sizelum} may arise from the
higher total star formation rates of the high redshift galaxies.

For the majority of the samples, an even stronger relation arises if we
compare the clump $\Sigma_{\rmn{SFR}}$ to the galaxy-averaged
$\Sigma_{\rmn{SFR}}$.  This is shown in the right-hand panel of Figure
\ref{fig:clumpplots}, and has a correlation coeficient $\rho = 0.79$
with 6.6\,$\sigma$ significance. The ratio of clump-to-average
$\Sigma_{\rmn{SFR}}$ can be thought of as a measure of the `clumpiness'
of the galaxy.

We conclude that the
properties of star-forming clumps in a galaxy are strongly dependent
on the global $\Sigma_{\rmn{SFR}}$ of the galaxy. Galaxies
with higher overall $\Sigma_{\rmn{SFR}}$ have higher clump surface
densities and are correpondingly offset in the clump size -- star formation
rate relation.  While this accounts for some of the differences seen
in Fig.~\ref{fig:sizelum}, it is also clear that there are more
bright clumps in the higher redshift galaxies. We quantify this below.

\subsection{H{\sc ii} region luminosity functions}
\label{sec:lf}

\begin{figure*}
\includegraphics[height=84mm, angle=90]{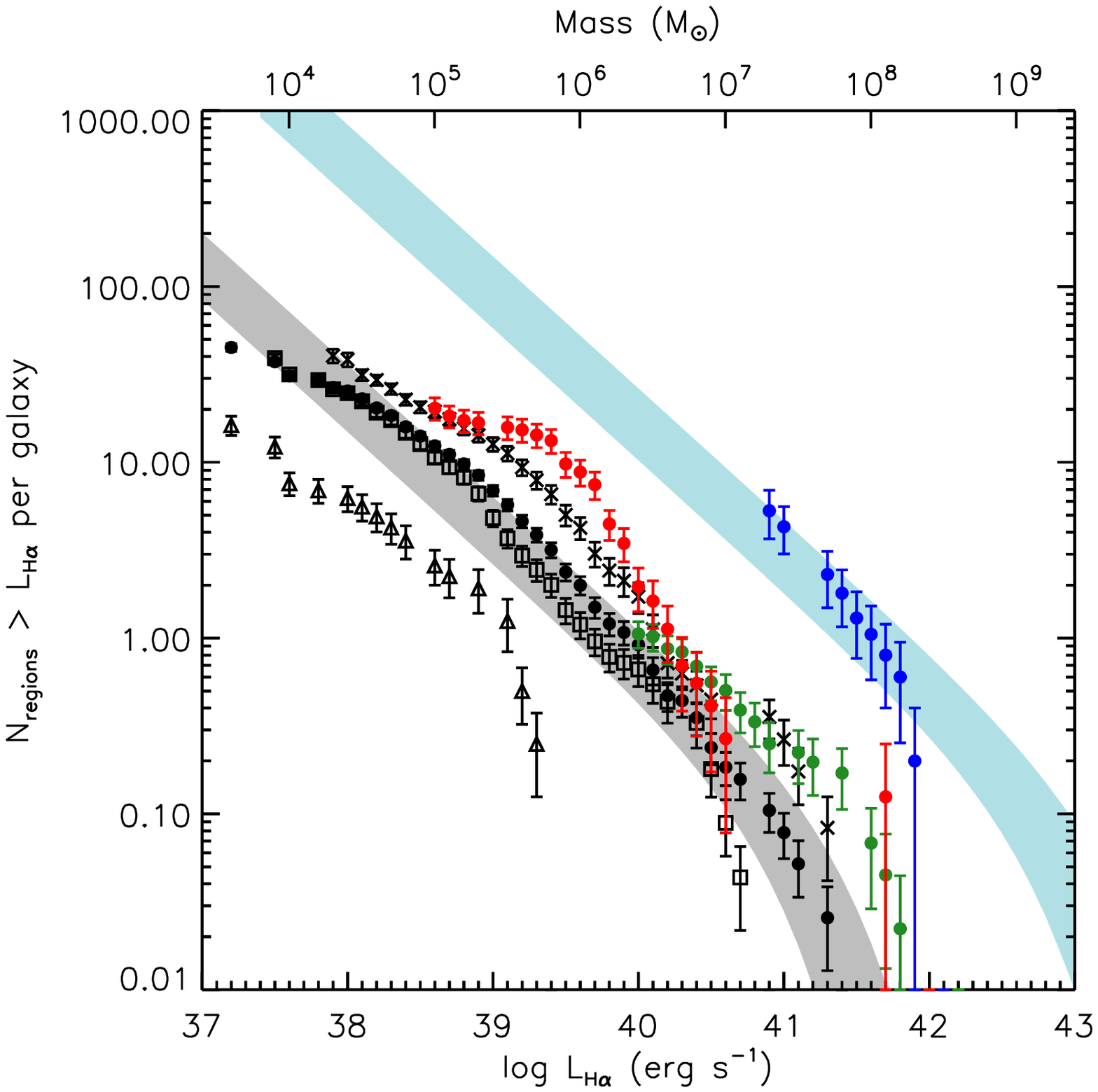}
\includegraphics[height=84mm, angle=90]{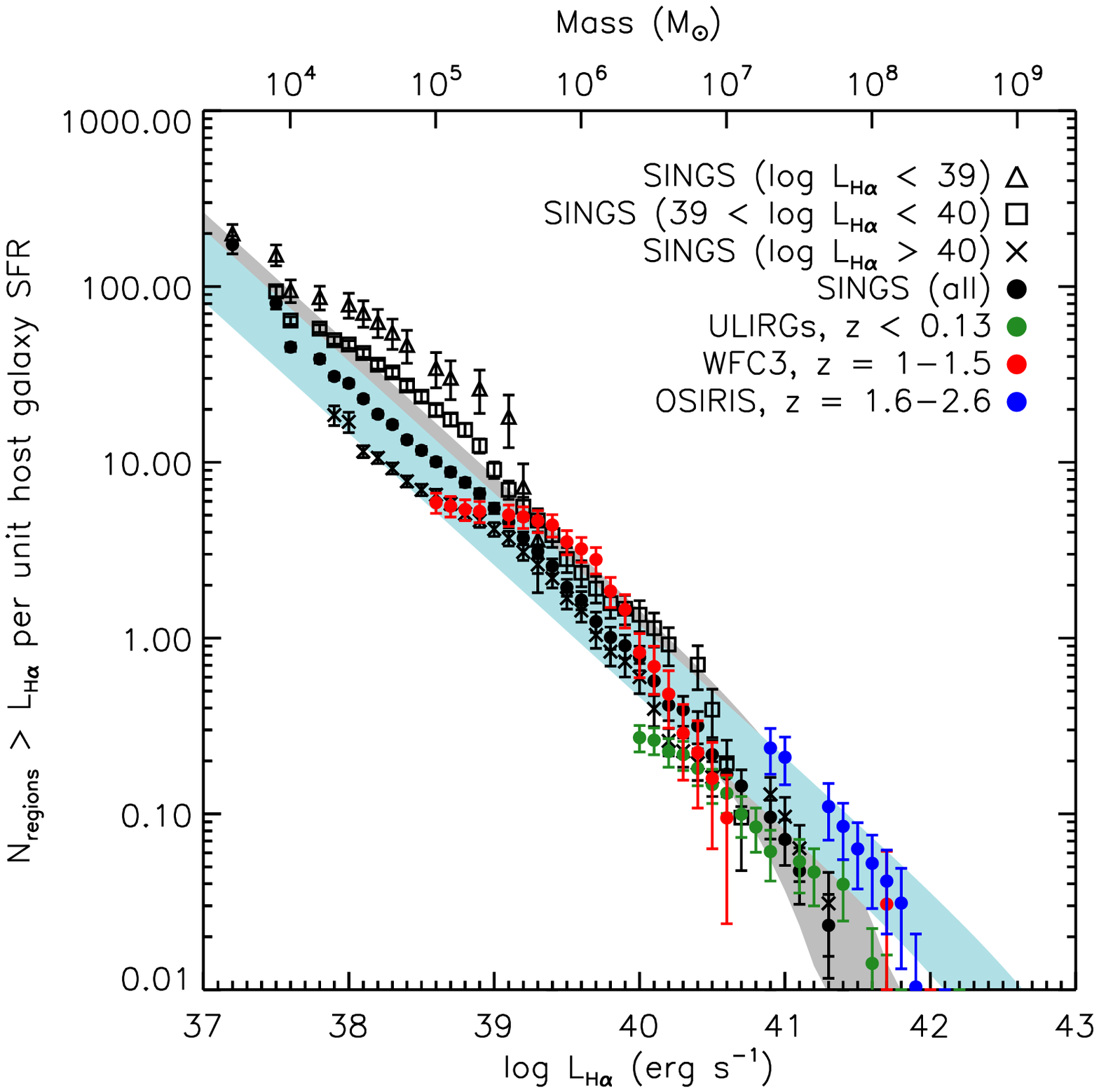}
\caption{Cumulative luminosity functions of H{\sc ii} regions in the SINGS, $z \sim 1 - 1.5$ and $z \sim 2$ samples, shown as a mean per galaxy (\emph{left}) and normalised by total galaxy H$\alpha$ luminosity (\emph{right}). We plot the SINGS sample as a whole and subdivded into three luminosity bins. The shaded regions illustrate model predictions from the GMC mass functions of \protect \citet{2012MNRAS.421.3488H} for Milky Way-like (grey) and high-$z$ (blue) simulations. We find evolution in the H{\sc ii} region luminosity function with redshift, which seems to be driven by the gas fraction of the disk.}
\label{fig:lf}
\end{figure*}

A quantitative measure of the clump brightness is to construct a
luminosity function (LF) of H{\sc ii} regions.  In the local universe,
the H{\sc ii} LF is presented in
\citet{1989ApJ...337..761K} and \citet{1997ApJS..108..199G}.  They demonstrate that
the LF can be fitted by a broken power-law, or by a power-law with an
exponential break.  In order to be consistent with our definitions of
clump sizes, we reanalyse the local data in order to construct our own
LF. The results are shown in Figure~\ref{fig:lf}. The left panel of the
figure shows the cumulative number of regions per galaxy as a function
of H$\alpha$ luminosity.  The normalisation of each bin takes into
account the different surface brightness limit of the galaxies, with
error bars computed from the Poisson error in counting regions. The
slope of the power-law part of the mass function is $\sim -0.75$, so
that although the LF appears steep in this representation, most of the
total luminosity is contributed by the brightest H{\sc ii} regions.

Solid black points show the average of all SINGS galaxies. However,
since we will be comparing the galaxies covering a range of
luminosities and redshifts, we have separated the galaxies from the SINGS
sample into 3 total H$\alpha$ luminosity bins. 
At a fixed luminosity, galaxies with lower total emission have fewer
regions, but the shape of the luminosity function is similar.
In order to emphasise the similarity of the mass function, we
normalise each of the curves by the total star formation rate of the host galaxies, and
show this in the right-hand panel. The similarity of all the H{\sc ii}
region LFs is now clear.  

There is a striking similarity between the LF of the $z \sim 1 - 1.5$ sample
and that of the highest SFR galaxies in the low-$z$ SINGS sample.  The
excess of very bright regions ($L\sim10^{41}\, {\rm erg}\,{\rm
s}^{-1}$) is down to one galaxy, Abell~773, which is the same compact
galaxy for which we found the clump surface brightnesses to be more
typical of the highest redshift galaxies.  The low luminosity slope of
the LF tends also to be flatter than that seen in the low redshift
galaxies, but it is hard to quantify this difference without directly
comparable surface brightness limits and is likely to be affected by
unresolved regions which are excluded. In any case, these faint regions
contribute little to the total flux.

In both panels, the H{\sc ii} region LF for the highest redshift
galaxies is strongly offset from the relation seen in the low redshift
SINGS sample and from the sample at $z \sim 1 - 1.5$, but is similar to the
low-redshift ULIRGs. Although the data do not probe the low-luminosity
slope of the LF, these galaxies have much brighter regions than are seen
at lower redshift. The right-hand panel emphasises that the this is not
because they contain many more regions overall.

In order to compare our data to models of mass functions, we must
relate the measured H$\alpha$ luminosities to model clump mass, $M$. As
an estimate, we use the H$\alpha$-derived SFR and adopt SFR
$\left(M_{\sun}\rmn{yr}^{-1}\right) = 4.6 \pm 2.6 \times 10^{-8}
M_{\sun}$ \citep{2010ApJ...724..687L}. This empirical relation is
based on local molecular clouds and applies to the high-density gas
where $A_K > 0.8$\,mag. However, we note that this relation is consistent with the
far-infrared-derived star formation rate and CO-derived gas masses of star-forming clumps reported in a lensed $z =
2.3$ galaxy by \mbox{\citet{2011ApJ...742...11S}}, but
clearly more high-resolution CO observations of high-$z$ galaxies are
required to confirm this. As a guide, we include this conversion on
the upper axis of Figure~\ref{fig:lf}.

The shapes of the LFs can be approximated by a power law with an
exponential cut-off at some high luminosity or mass. The difference
between the samples' LFs is then best described by a pure luminosity
evolution, so that the cut-off shifts to higher luminosity/mass at
higher redshift.

To demonstrate this, we include shaded regions representing a Schechter function of the form 

\begin{equation}
N\left( > M \right) = N_0 \left( \frac{M}{M_0} \right)^{\alpha}\exp\left(\frac{-M}{M_0}\right),
\end{equation}

where we adopt the median value of $\alpha = -0.75$ from
\citet{2012MNRAS.421.3488H}. The normalisation $N_0$ is arbitrary, so
we fit $N_0$ to the $z=0$ data and then keep it fixed while allowing
$M_0$ to vary in order to find best-fit values of in the different
samples. The best-fit values are $M_0 = 4.6^{+3.1}_{-2.0} \times
10^7\,M_{\sun}$ at $z=0$, $M_0 = 8.0^{+11.0}_{-4.3} \times
10^7\,M_{\sun}$ at $z \sim 1 - 1.5$ and $M_0 = 1.5^{+2.2}_{-1.0} \times
10^9\,M_{\sun}$ at $z \sim 2$, where the errors are found with a
bootstrap method. We shade the best-fit Schechter functions at $z = 0$
and $z \sim 2$ in grey and blue respectively in
Figure~\ref{fig:lf}. The normalisation of the model in the right-hand
panel is obtaines by summing the luminosities of the clumps. The result
is in remarkably good agreement with our observations.

Not only do the highest-redshift galaxies have H{\sc ii} regions that
are higher surface brightness, but the characteristic luminosity of the
regions is higher too.  We suggest that the presence of high-luminosity
regions may be a good operational definition of the clumpiness of a galaxy.

\section{Discussion}
\label{sec:disc}

In the previous section, we presented an analysis of star forming
regions in galaxies at $z=0$, $z\sim 1 - 1.5$ and $z \sim 2$.  We find that
the luminosities of the regions in $z=1$ galaxies are similar to those of
bright ($L > 10^{40}$\,erg\,s$^{-1}$) galaxies at low
redshift, but the surface brightnesses are systematically higher.  At higher
redshifts, the properties of the galaxies change, with the galaxies
having clumps that are both much higher surface brightness and shifted to
higher total luminosities.  This accounts for the qualitative
impression that the most distant galaxies are ``clumpier.''

We also noted that the increase in the surface brightness of H{\sc ii}
regions tracks the increase in the average star formation rate
surface density, $\Sigma_{SFR}$, of the galaxies. The observations are
consistent with the changing properties of the H{\sc ii} regions being
driven by changes in the overall $\Sigma_{SFR}$ of the galaxies. 
We can link the increase in the observed $\Sigma_{SFR}$ to an
increase in the gas surface density of these galaxies by assuming that the
Kennicutt-Schmidt law holds at $z \sim 2$ as well as at $z=0$. In this
case, we have an emerging picture that the changes we see are likely driven
by greater gas surface densities at higher redshift.

The connection between the increasing surface density of clumps and the
greater peak brightness arises naturally from this picture
\citep{2011arXiv1111.2863H}. The clump mass required for collapse on
scale $R$ from a turbulent ISM is given by the Jean's mass, $M_J$:
\begin{equation}
\rho_c = \frac{3}{4\pi R^3}M_J \approx \frac{9}{8\pi R^2 G}\sigma_t(R)^2,
\label{eq:dencrit}
\end{equation}
where $\sigma_t(R)$ is the line of sight turbulent velocity
dispersion. Assuming a turbulent velocity power spectrum $E(k)$, the velocity dispersion $\sigma_t(R)$ for wavenumber $k=1/R$ is
\begin{equation}
\sigma_t^2 = k E(k) \propto R^{p-1},
\label{eq:sigma_t}
\end{equation}
where $p\approx2$ for supersonic turbulence appropriate to the ISM \citep{burgers74,2009A&A...494..127S}.

To determine the normalisation of the relation, we assume that the
clumps are located in a marginally unstable disk. We note that the available kinematic data for the $z \sim 2$ sample and for MACS~J1149 support this assumption, as do larger surveys \mbox{\citep{2011ApJ...733..101G}}; nonetheless, clearly this is uncertain without dynamical data for the entire sample. However, if we make the assumption that the galaxies are rotating disks with Toomre parameter $Q \approx 1$, we can relate the epicyclic frequency, $\kappa$ of the disk to its scale
height, $h$:
$\kappa \approx \sigma_t(h)/h$.
\begin{equation}
Q = \frac{\kappa \sigma_t(h)}{\pi G \Sigma_0} \approx \frac{\sigma_t(h)^2}{\pi G \Sigma_0 h}. 
\label{eq:toomre}
\end{equation}
Since the stability of the disk is a global phenomenon, we will
associate $\Sigma_0$ with the average surface density of the star
forming disk, $\Sigma_{\rm disk}$, and treat the quantities appearing in Eq.~\ref{eq:toomre}
as appropriate global disk averages. Since the disk is made up of both stars and gas,
we must take an appropriate average of the surface densities in the two components.
Following \citet{2001MNRAS.323..445R}, and focussing on the largest unstable fluctuations,
the appropriate combination of gas and star surface densities (denoted $\Sigma_g$ and $\Sigma_*$)
is
\begin{equation}
\Sigma_{\rmn{disk}} = \Sigma_g  + \left(\frac{2}{1+f_{\sigma}^2}\right)\Sigma_{\ast}, 
\end{equation}
where $f_{\sigma} = \sigma_*/\sigma_t$ is the ratio of the velocity dispersion of the stellar component to that of the gas. 

Assuming $Q\approx 1$, we can write,
\begin{equation}
\sigma_t^2(R) = (\pi G \Sigma_{\rm disk} h) \left(\frac{R}{h}\right)^{p-1} 
        \approx \pi G \Sigma_{\rm disk} R,
\label{eq:sigma_t2}
\end{equation}
where we have used Eq.~\ref{eq:sigma_t} to relate $\sigma_t(h)$ to
$\sigma_t(R)$. Combining with Eq.~\ref{eq:dencrit} gives a critical
density for collapse of
\begin{equation}
\rho_c(R) = \frac{9}{8} \Sigma_{\rm disk} \frac{1}{R}.
\end{equation}
Assuming that the cloud contracts by a factor $\approx 2.5$ as it
collapses, the post-collapse surface density is
\begin{equation}
\Sigma_{\rm cloud} \approx 10 \rho_c R    
       \approx 10\, \Sigma_{\rm disk}.     
\label{eq:dencrit2}
\end{equation}
Thus, for the turbulent power spectrum $p\approx 2$, the surface
density of collapsed clouds is independent of radius and proportional
to the surface density of the disk.  The normalisation of the relation
follows from the collapse factor and the requirement that the disk is
marginally stable.  This model provides a good description of clouds in
the Milky Way (Larson's laws) as discussed in
\citet{2011arXiv1111.2863H}, and predicts that the surface brightnesses
of clouds should increase as the gas surface density (and thus overall
average star formation rate surface density) increases.  If we assume a
constant Kennicutt-Schmidt law of the form $\Sigma_{\rm gas} \propto
\Sigma_{\rm SFR}^{1.4}$ \citep{1998ApJ...498..541K}, we can compare
Eq.~\ref{eq:dencrit} to our data; we therefore overplot this line on
Figure~\ref{fig:clumpplots} and find that it is in good agreement with
the observations.

Moreover, the model predicts that the most massive clouds should
increase in size with the average gas surface density.  This follows from the
marginal stability condition (Eq.~\ref{eq:toomre}), since density
structures on scales greater than $h$ will tend to be stabilised by
disk rotation.

This can be demonstrated formally by examining the dispersion relation
for a finite thickness disk \citep{2009ApJ...702L...5B}. \citet{2011arXiv1111.2863H} shows that this leads to an exponential cut-off
of the clump mass function above a mass

\begin{equation}
M_0 \approx \frac{4\pi}{3} \rho_c(h)  h^3 = \frac{3\pi^3G^2}{2}\frac{\Sigma_{\rm disk}^3}{\kappa^4},
\end{equation}

where we have used Eqs.~\ref{eq:sigma_t} and \ref{eq:toomre} to
express $h$ as a function of $\Sigma_{\rm disk}$ and $\kappa = v_{\rm disk}
/R_{\rm disk}$ (where $v_{\rm disk}$ is the disk
circular velocity and $R_{\rm disk}$ is half-mass radius of the
disk). Expressing $\kappa$ in units of $\rm 100\,{\rm km}\,{\rm s}^{-1}\,kpc^{-1}$ we obtain a normalisation of

\begin{equation}
\frac{M_0}{M_{\odot}} = 8.6\times10^{3} \left( \frac{\Sigma_{\rm disk}}{10\,M_{\odot}\,{\rm pc}^{-2}}\right)^3
        \left( \frac{\kappa}{{\rm 100\,{\rm
                km/s}\,kpc^{-1}}}\right)^{-4}.
\label{eq:m0}
\end{equation}
We can check that this results in a reasonable value of $M_0$ in the Milky Way by using a gas surface density $\Sigma_{\rmn{gas}} \sim 10\,M_{\sun}\,{\rm pc}^{-2}$ and a gas fraction of 10\% with $f_{\sigma} = 2$ \citep{2003AJ....126.2896K} to obtain an effective $\Sigma_{\rmn{disk}} \sim 35\,M_{\sun}\,{\rm pc}^{-2}$. With $\kappa = 36.7\,{\rm km}\,{\rm s}^{-1}\,{\rm kpc}^{-1}$ \citep{1997MNRAS.291..683F}, this gives $M_0 \sim 10^7 M_{\sun}$, in good agreement with the characteristic mass of the largest Galactic GMCs \citep{2006ApJ...641L.113S}. 

\begin{figure*}
\includegraphics[height=184mm, angle=90]{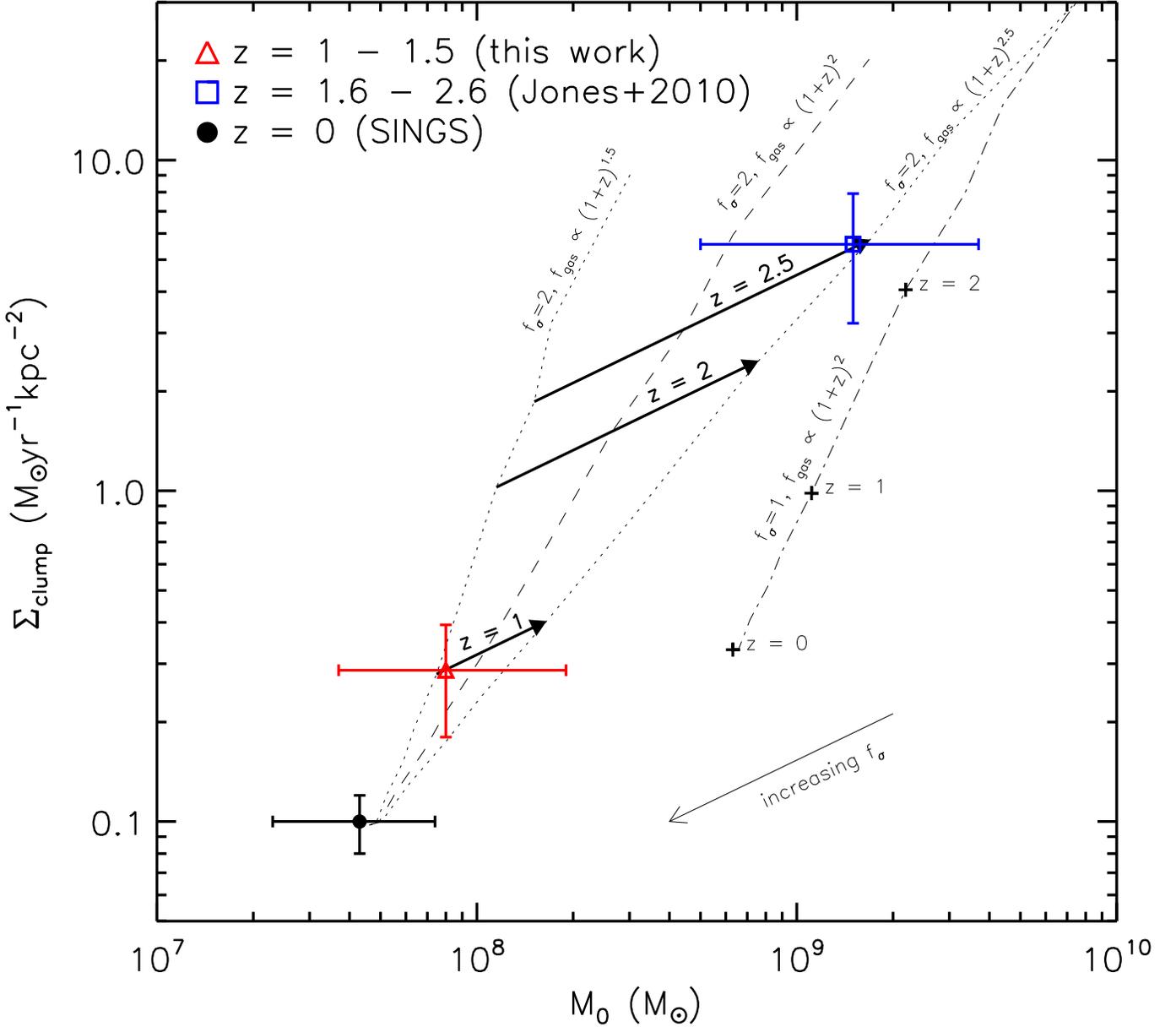}
\caption{The evolution of cut-off mass $M_0$ and clump surface density $\Sigma_{\rmn{clump}}$ in comparison with model predictions (equation \ref{eq:m0}). The model is strongly dependent on the assumed evolution of gas fraction with redshift; we show tracks for $f_{\rmn{gas}} \propto (1+z)^{2 \pm 0.5}$ and the thick arrows show the effect of increasing the gas fraction. We also show a track for $f_{\sigma} = 1$ - i.e. assuming that the gas and stars have the same velocity dispersion. This makes the disk unstable on larger scales and would lead to higher-mass clumps at low-$z$ than are observed. The impact of this is reduced at higher-$z$ where gas dominates the disk dynamics. As $f_{\sigma}$ increases, the points move in the direction of the arrow in the lower right corner. The model provides a good fit to the data, demonstrating that larger, higher surface brightness clumps at high-$z$ are a natural consequence of increasing gas fractions, which explains the observed morphologies of high-$z$ galaxies.}
\label{fig:m0sig}
\end{figure*}

Eq.~\ref{eq:m0} shows that the mass cut-off depends strongly on the disk
surface density --- the higher the surface density, the more massive
the clumps that are able to form. This trend can, however, be opposed by the
stablising effect of angular rotational speed. For a fixed disk
radius, a higher circular velocity tends to reduce the mass of the
largest clumps.

For low-redshift galaxies, simple theoretical models suggest that
$R_{disk} \propto v_{\rm disk}$ since halo spin is weakly dependent
on the halo mass \citep{1998MNRAS.295..319M}, and thus we should expect
the dependence on the disk surface density to be the dominant trend
controlling the cut-off clump mass. This is confirmed by analysis of the
observed properties of galaxies. For example, \citet{2011MNRAS.410.1660D} find
\begin{equation}
\frac{R_{\rm disk}}{\rm kpc} \approx 2.5 \left( \frac{v_{\rm
      disk}}{100\,{\rm km/s}} \right)^{1.2}
\end{equation}
in the local universe. Combining this with the observed dependence of the
disk rotation velocity on galaxy mass
\begin{equation}
\frac{M_*}{10^{10} M_{\sun}}
  = 0.25 \left(\frac{v_{\rm disk}}{100\,{\rm km/s}}\right)^{4.5},
\end{equation}
we obtain
\begin{equation}
\frac{\kappa}{\rm 100\,{\rm km/s}\,kpc^{-1}}
  \approx 0.38 \left(\frac{M_*}{10^{10} M_{\sun}}\right)^{-0.04},
\end{equation}
which shows that $\kappa$ is very weakly dependent on the
galaxy mass, and the variation in the clump mass functions of
local galaxies is driven is driven by the disk surface density.

We also note that in a disk of constant circular velocity, $\kappa$ scales with radial distance $r$ as $\kappa \propto r^{-1}$, while the gas surface density profiles are shallow, with $\Sigma_{\rm gas} \propto r^{-4/3}$ \citep{2010MNRAS.409..515F}. Thus from Eq.~\ref{eq:m0} there is no dependence of $M_0$ on $r$; while the surface density is higher towards the centre of the galaxy, this is balanced by the higher rotational frequency. This explains the observations that clump properties appear to be driven by the global properties of their host galaxies rather than by local conditions, and this allows us to use disk-averaged values of $\kappa$ and $\Sigma$.

We have no measurements of the gas contents of our samples, but
dynamical information available for the $z \sim 2$ sample permits us to
predict the cut-off mass in these galaxies from our model if we
estimate $\Sigma_{\rmn{disk}}$ from the dynamical mass. Using the
measurements reported in \citet[see their Table
2]{2010MNRAS.404.1247J}, we compute a cut-off mass ranging from $3.3
\times 10^6 - 3.1 \times 10^9\,M_{\sun}$ for the $z \sim 2$ sample. The
median value is $5 \times 10^7\,M_{\sun}$, approximately $5\times$
higher than the Milky Way value. We therefore expect that the $z \sim
2$ sample should contain clumps of higher mass and
luminosity, as observed. However, the uncertainty in the cut-off mass
for the $z \sim 2$ sample is very large due to uncertainties in
$\Sigma_{\rmn{disk}}$ and $\kappa$, which prevents us from making a
precise comparison of the cut-off mass in the different samples.

To understand how the clumpiness of galaxies evolves, we must therefore use simulations to estimate the evolution of their scaling relations. \citet{2011MNRAS.410.1660D} present a simple analytic model that seems to describe the observational data well \citep{2006ApJ...650...18T,2009ApJ...706.1364F,2010ApJ...713..738W}. We use their scaling relations for mass, size and rotational frequency with redshift in combination with Eq. \ref{eq:dencrit2} and Eq. \ref{eq:m0} to predict how the cut-off mass and clump surface brightness should evolve with redshift. Figure \ref{fig:m0sig} illustrates this evolution for a gas fraction evolution of $f_{\rmn{gas}} \propto \left( 1 + z \right)^{2\pm 0.5}$ \citep{2011ApJ...730L..19G}. The arrows show how altering the assumed gas fraction changes the model. This suggests that the changing clump properties are a natural consequence of increasing gas fractions dominating high-$z$ galaxy dynamics.  The high gas fractions probably arise from high gas infall rates at high redshift \citep{2009ApJ...694L.158B,2010MNRAS.406..112K,2011ApJ...730....4B}; however, our observations do not directly rule in or out cold flows. Our results merely require high gas fractions, and cold flows are a method of maintaining the gas supply. Crucially, we note that this effect is tempered by the more compact nature of galaxies, which leads to higher epicyclic frequencies that limit the collapse on larger scales. The need to include the $\kappa$ term is apparent from our H{\sc ii} region luminosity functions: without it, a factor 10 increase in disk surface density would correspond to an increase in clump luminosity of $1000\times$, and we do not observe such a large increase.

To summarise, we find that our simple theoretical model is in good agreement with the observations and suggests that the evolving `clumpiness' of galaxies is a manifestation of the different cut-off mass of the H{\sc ii} region luminosity function, which is driven by evolution in the gas fraction with redshift.

\section{Conclusions}
\label{sec:conc}

We have used \emph{HST/WFC3} to obtain narrowband H$\alpha$ imaging of eight gravitationally lensed galaxies at $z \sim 1-1.5$. The magnification provided by the lensing enables us to reach spatial resolutions in the source plane of $68 - 615$\,pc. In addition, to provide comparisons we have re-analysed the lensed $z \sim 2$ sample observed with Keck/OSIRIS by \citet{2010MNRAS.404.1247J}, the \citet{2011A&A...527A..60R} sample of $z < 0.13$ (U)LIRGs observed with VLT/VIMOS and the H$\alpha$ narrowband imaging of the $z = 0$ SINGS survey \citep{2003PASP..115..928K}.

The high-$z$ samples have `clumpy' morphologies, dominated by a few large regions of high H$\alpha$ luminosity, which we use as a proxy for the SFR. We have extracted star-forming clumps from the galaxies in each sample and examined their properties. The clumps follow similar SFR-size scaling relations in all samples, but the normalisation of the relation exhibits systematic offsets to higher surface brightness at higher redshifts. The normalisation appears to be approximately constant within a given galaxy, implying that this relation is driven by global galaxy properties.

On comparison with the properties of the host galaxies, we find that all samples follow approximately the same scaling relations between the clump surface brightness and both the host galaxy's total H$\alpha$ luminosity, $L_{{\rm H}\alpha}$, and its average surface density of star formation, $\Sigma_{\rmn{SFR}}$, and that they evolve along this relation in decreasing $L_{{\rm H}\alpha}$ and $\Sigma_{\rmn{SFR}}$ with decreasing redshift.

We have measured the luminosity function of clumps in the samples, and shown that the $z \sim 1 - 1.5$ sample is similar to the higher-$L_{{\rm H}\alpha}$ members of the SINGS sample. When normalised by the host galaxies' total SFR, the SINGS and $z \sim 1 - 1.5$ samples can be fit by the same Schechter function, while the ULIRGs and $z \sim 2$ samples are offset horizontally. This shift can be explained by an increase in the cut-off mass of the H{\sc ii} region luminosity functions of the ULIRGs and $z \sim 2$ disks.

We present a simple theoretical model which shows that the evolution in luminosity and surface brightness are connected, and are driven by the competing effects of disk surface density $\Sigma_{\rm disk}$ and the epicyclic frequency $\kappa$. Galaxies at high redshift tend to have higher $\Sigma_{\rmn{disk}}$, which increases the maximum mass of clumps that are able to form; however, this is tempered by the more compact nature of high-$z$ galaxies, implying higher $\kappa$, which impedes collapse on the largest scales.

We have shown that this model is consistent with the evolution in clump properties seen in our data. We therefore conclude that the clumps observed in high-$z$ galaxies are star-forming regions analogous to those found locally but with higher masses and surface brightnesses. As H{\sc ii} regions in the distant Universe are larger and brighter, they give rise to the `clumpy' appearance. The increase in clump luminosity is driven primarily by increasing gas fractions at high-$z$. This clearly motivates further study with ALMA to better quantify the evolution of gas properties in high-$z$ galaxies.

\section*{Acknowledgments}

The authors would like to thank Karl Glazebrook, Emily Wisnioski, Lisa Kewley and Norm Murray for useful discussions and Andrew Newman for providing an updated strong lensing model of the cluster Abell 611. RCL acknowledges a studentship from STFC, RGB and IRS are supported by STFC and IRS further acknowledges a Leverhulme Senior Fellowship. AMS acknowledges an STFC Advanced Fellowship, and JR is supported by the Marie Curie Career Integration Grant 294074. HE gratefully acknowledges financial support from STScI grants GO-09722, GO-10491, GO-10875, and GO-12166. This work is based on observations with the NASA/ESA Hubble Space Telescope obtained at the Space Telescope Science Institute, which is operated by the Association of Universities for Research in Astronomy, Incorporated, under NASA contract NAS5-26555. Support for Program number 12197 and Program number 11678 was provided by NASA through a grant from the Space Telescope Science Institute, which is operated by the Association of Universities for Research in Astronomy, Incorporated, under NASA contract NAS5-26555.

\bibliographystyle{mn2e}
\bibliography{bib}

\bsp

\label{lastpage}

\end{document}